\documentstyle[12pt,epsfig,psfig]{article}
\begin{document}

\newcommand{\beq}{\begin{equation}}
\newcommand{\eeq}{\end{equation}}
\newcommand{\nn}{\nonumber}

\def\ii{\'{\char'20}}
\def\r{\rightarrow}
\def\err{\end{array}}
\def\bea{\begin{eqnarray}}
\def\eea{\end{eqnarray}}
\def\bp{{\bf p}}
\def\bk{{\bf k}}
\def\bq{{\bf q}}
\def\ttau{\tilde{\tau}}
\def\tchi{\tilde{\chi}}
\def\trho{\tilde{\rho}}
\def\teps{\tilde{\epsilon}}
\def\tnu{\tilde{\nu}}
\def\tgamma{\tilde{\gamma}}

\title{Instanton propagator and instanton induced processes 
in scalar model}

\author{ Yu.A. Kubyshin\footnote{E-mail: \tt kubyshin@theory.npi.msu.su} \\
{\small \em Institute for Nuclear Physics, Moscow State University}  \\
{\small \em 117899 Moscow, Russia \vspace{0.2cm} }\\ \vspace{0.2cm}
and \\ 
P.G. Tinyakov \\
{\small \em Institute for Nuclear Research of the Russian Academy of 
Sciences}\\
{\small \em 60th October Anniversary prospect, 7a 
117312 Moscow, Russia}\\ 
{\small \em and Institute of Theoretical Physics, University of Lausanne}\\
{\small \em BSP 1015 Dorigny, Switzerland}
} 

\date{}

\maketitle

\begin{abstract} 
The propagator in the instanton background in the
$(- \lambda \phi^{4})$ scalar model in four dimensions is studied.
Leading and sub-leading terms of its asymptotics for large momenta
and its on-shell double residue are calculated analytically. 
These results are
applied to the analysis of the initial-state and initial-final-state
corrections and the calculation of the next-to-leading (propagator)
correction to the exponent of the cross section of instanton induced 
multiparticle scattering processes.
\end{abstract}

\section{Introduction}
\label{introduction}

There is a number of interesting physical effects induced by instanton
solutions. They appear in theories which possess a non-trivial
structure of vacua. The most prominent example is the electroweak
theory with an infinite number of vacuum states labelled by the
Chern-Simons number \cite{EWT-vac}. The instanton solutions describe
transitions with baryon number violation between the vacua. Another
example of the instanton induced process is the decay of a metastable
(false) vacuum due to underbarrier tunelling from a false vacuum to
the true one \cite{VolKobOkun}.  The third example is a shadow process
\cite{shadow}. This is a non-perturbative process in which both the
initial and the final state are in the false vacuum.  Apart from
standard perturbative contributions, the processes which start and end
in the false vacuum acquire additional contributions due to the
underbarrier tunelling of the system to another vacuum and its return
to the initial one. This transition is obviously induced by an
instanton solution and goes through the intermediate state containing
a bubble of the true vacuum. In this article we study the cross section 
of shadow processes.  

Much work has been done to study of the instanton induced transitions,
and quite effective techniques for the calculation of the
probabilities of such transitions have been developed (see Refs.
\cite{Ti1,Mat} for a review). Consider for simplicity a scalar theory
with the field $\phi(x)$ and the action $S(\phi)$. Let us discuss a
process ($2 \; \rightarrow$ any) with two initial particles of the
total energy $E$. The total cross section of this process is given by
the sum over the partial cross sections, 
\[
\sigma_{2}(E) = \sum_{n}
\sigma_{2 \rightarrow n} (E),
\]
and each term is calculated from the amplitude
${\cal A}_{2 \rightarrow n}(p_{1},p_{2};q_{1}, \ldots , q_{n})$
in a standard way. The amplitude is obtained by applying the
Lemann-Symanzyk-Zimmermann reduction formulas to the $(n+2)$-point
Green function,
\beq
G_{n+2}(x_{1}, \ldots , x_{n+2}) = \int D\phi(x) e^{-S(\phi)}
\phi(x_{1}) \ldots \phi(x_{n+2}).   \label{GF-1}
\eeq
If the model possesses an instanton solution $\phi_{inst}(x)$, then
besides the standard perturbative contribution to the Green function,
corresponding to the expansion of Eq.~(\ref{GF-1}) around the trivial
solution $\phi=0$, there exists a contribution due to the instanton
sector. It is the contribution
we will focus on in this paper. It can be calculated by
performing the expansion $\phi(x) = \phi_{inst}(x) + \eta(x)$
and integrating over the fluctuations $\eta(x)$ around the instanton:
\bea
G_{n+2}(x_{1}, \ldots , x_{n+2}) & = & \int D\eta(x) e^{-S_{inst}}
\exp \left\{- \frac{1}{2} \int dx \eta(x) \hat{D}_{x} \eta(x) -
\ldots \right\}   \nonumber \\
& \times & \left[ \phi_{inst}(x_{1}) + \eta(x_{1}) \right] \ldots
\left[ \phi_{inst}(x_{n+2}) + \eta(x_{n+2}) \right].
\label{GF-2}
\eea
Here $S_{inst}$ is the instanton action and $\hat{D}_{x}$ is the
second order differential operator determined by the quadratic in
$\eta(x)$ part of the action.
The dots in the exponent stand for cubic and higher order terms in
$\eta(x)$.
In general the instanton solution is parametrized by a
number of continuos parameters denoted here by $\zeta_{A}$. 
These parameters correspond to symmetries of the model.
Thus, in fact, there is an infinite family of instanton
configurations which have to be taken into account. Due to this fact
the operator  $\hat{D}_{x}$ possesses zero modes associated with
these symmetries, and is not invertible. Functional integration over the
fluctuations $\eta (x)$ is perfomed according to a known 
procedure, and, as a result, instanton contribution (\ref{GF-2}) to the 
Green function $G_{n+2}(x_{1}, \ldots , x_{n+2})$ includes the 
integral over the parameters $\zeta_{A}$ (see ref.\cite{Ti1} for details).
In the leading semiclassical approximation the result can be 
schematically written as 
\beq
G_{n+2}(x_{1}, \ldots , x_{n+2}) \sim \int d\zeta_{A} e^{-S_{inst}}
 \phi_{inst}(x_{1};\zeta) \ldots \phi_{inst}(x_{n+2};\zeta) + 
\mbox{corrections}.
\label{GF-3}
\eeq

Using Eq.~(\ref{GF-3}) one can evaluate the partial cross-sections
$\sigma_{2 \rightarrow n}(E)$, and after performing the summation
over $n$ one obtains the leading-order semiclassical expression for
the total cross-section:
\beq
\sigma_{2}(E) \sim \exp \left\{\frac{1}{\lambda}
\left( F^{(L)}(E/E_{sph}) + \ldots \right) +
{\cal O}(1) \right\},
         \label{sigma-LO}
\eeq
where $\lambda$ is the coupling constant in the model and
$E_{sph}$ is the energy of the sphaleron configuration which
characterizes the height of the barrier separating the vacua.
The superscript $(L)$ stands for "leading".
On dimensional grounds it can be shown that
\beq
E_{sph} = \kappa \frac{m}{\lambda},    \label{E-sph}
\eeq
where $\kappa$ is a numerical constant depending on the model.
It is easy to see that for a shadow process 
\[
F^{(L)}(0) = 2\lambda S_{inst}.
\]
The dots in the exponent of the cross section in Eq. (\ref{sigma-LO}) 
stand for $\lambda^{0}$ terms. 
In the weak coupling regime, $\lambda \rightarrow 0$, contributions
described by this function dominate and the
${\cal O}(1)$ term in Eq. (\ref{sigma-LO}) gives an exponentially
small correction.

The first calculations of instanton induced processes were done for 
the electroweak theory in Refs. \cite{Ri1}.  
There the role of $\lambda$ is played by $g^{2}$,
the square of the gauge coupling constant and $S_{inst} = 8\pi^{2}/g^2$. 
The leading order contribution to the function $F(E/E_{sph})$ is 
equal to 
\[
F^{(L)} \left(\frac{E}{E_{0}} \right) =
- 16 \pi^{2}
 \left[ 1 - \frac{9}{8} \left( \frac{E}{E_{0}} \right)^{4/3} \right]
,
\]
where the parameter $E_{0} = \sqrt{6} M_{W}/\alpha_{W}$ is of the 
order of the sphaleron energy $E_{sph} \approx 10$TeV  \cite{KRT,ew-LO}

In the scalar $(- \lambda \phi^{4})$-theory, which we are going to
consider in this paper, the leading contribution is equal to
\beq F^{(L)}\left(\epsilon \right) = - 32 \pi^{2} + \frac{2
  \kappa \epsilon}{\ln^{1/2} (1/\epsilon) },
\label{F-LO}
\eeq
where $\epsilon = E/E_{sph}$ and $E_{sph}$ is given by Eq.
(\ref{E-sph}) with $\kappa = 113.4$ \cite{Ti1,KT}.

An effective method of calculation of the corrections to the 
function $F(E/E_{sph})$ was proposed and developed in Refs. \cite{KRT}. 

{}From the  above examples it is clear that at energies
$E \ll E_{sph}$ the cross section of the instanton induced process
is exponentially supressed. On the other hand there are reasons
to expect that as the total energy of the initial states grows, 
the probability of the underbarrier transition, induced by the instanton, 
grows as well and the supression becomes weaker. For high energies 
the function $F(E/E_{sph})$ may be close enough to zero and the 
instanton induced processes may become observable.
This opens a possibility of new interesting physics within the 
Standard Model at future colliders \cite{Ri2}.  
However, in Ref. \cite{Rub} arguments were presented which suggest  
that perhaps the function $F(E/E_{sph})$ never vanishes and, hence, 
at all energies some supression always remains. 

The considerations above are based on the assumption that
the complete expression for the total cross section, not just
its leading order contribution, can be presented in the
exponential form, i.e.
\beq
\sigma_{2}(E) \sim e^{\frac{1}{\lambda} F(\epsilon) +
{\cal O}(1) }.
         \label{sigma2}
\eeq
The leading order term in $F$ has been just discussed. The
next-to-leading term is a propagator correction, for after 
integration over $\eta(x)$ in Eq.~(\ref{GF-2}) it includes
contributions from the propagator in the instanton background. 
We will call this propagator the instanton propagator for shortness. 
It is, of course, equal to $\hat{D}_{x}^{-1}$ restricted to the space of 
fields orthogonal to the zero modes. In fact, as we will see, 
calculation of the next-to-leading correction requires knowledge
of the exact expression for the residue of the instanton propagator. 
It turns out that in the
$(-\lambda \phi^{4})$-theory an exact expression for the
instanton propagator can be obtained.
Derivation of the exact formula for the residue of the instanton 
propagator, as well as calculation and discussion of the
propagator correction to the function $F(\epsilon)$ in the scalar model
is one of the purposes of the present paper.

An important issue in the theory of instanton induced processes 
is the validity of formula (\ref{sigma2}). The difficulty in proving it 
is due to the initial state corrections and initial-final 
state corrections. In general, it is not easy to show that 
they exponentiate, so that the whole result can be represented in the form 
given by Eq. (\ref{sigma2}). In the
electroweak theory a proof, based on the properties of the propagator
in the instanton background was given in Ref. \cite{Mu92}. We
apply the arguments of Ref. \cite{Mu92} in the $(-\lambda \phi^{4})$-theory
making use of the explicit expression for the propagator.

On the other hand, it was shown (see Refs. \cite{RT,Ti2,RST}) that for
the ($N \rightarrow \mbox{any}$) process, where $N \sim 1/\lambda$
and, therefore, is parametrically large for small $\lambda$, the total
transition probability in the semiclassical approximation is given by
\beq
\sigma_{N}(E) \sim e^{\frac{1}{\lambda} F(\epsilon, \nu) +
{\cal O}(1) }.    \label{sigma-N}
\eeq
Here we introduced the parameter $\nu = N/N_{sph}$, where 
\beq
  N_{sph} = \frac{\kappa'}{\lambda}   \label{N-sph}
\eeq
is the characteristic number of particles contained in the sphaleron. 
In Refs. \cite{RT,Ti2} it was conjectured that in the leading 
semiclassical approximation the two-particle cross section can be
calculated from the following formula:
\beq
\lim_{\lambda \rightarrow 0} \lambda \ln \sigma_{2} =
\lim_{\nu \rightarrow 0} F \left( \epsilon, \nu \right),
\label{conj}
\eeq provided the limit $\nu \rightarrow 0$ exists. The problem with
this conjecture is that the function $F(\epsilon, \nu)$ is known to
contain contributions singular in $\nu$. In particular, such
contributions appear in the course of the calculation of the
propagator correction. The conjecture of Refs. \cite{RT,Ti2} 
basically claims that all such
terms cancel each other in the final answer.  Its validity, of course,
means that the semiclassical form of the two-particle cross section is
indeed given by Eq. (\ref{sigma2}) with $F(\epsilon) = F(\epsilon,0)$.  
Verification of conjecture (\ref{conj}) in the next-to-leading
order is another purpose of this paper. Note that different arguments
in favor of this conjecture were given in
Refs. \cite{Mueller,RubReb}.

The plan of the article is the following. 
We start Sect.~\ref{PT} with a review of basic facts 
about the perturbation theory around the instanton following mainly 
the results of Refs. \cite{KRT}, \cite{Ti2}. We also discuss the 
general structure of the leading and next-to-leading corrections to the 
function $F(\epsilon,\nu)$ and saddle point equations defining these 
corrections in the semiclassical approximation. Finally, using a 
formula for the leading asymptotics of the instanton propagator, 
obtained in Ref. \cite{Vo1}, we explain the appearance of terms 
singular in $\nu$ as $\nu \rightarrow 0$ and prove that they cancel each 
other. In Sect.~\ref{model} we
describe the model and review the derivation of the exact explicit
expression for the instanton propagator obtained in this model 
some time ago in Ref. \cite{Ku}. In Sect.~\ref{asympt} we
calculate three leading terms of the large-$s$ asymptotics of
the Fourier transform of the propagator and discuss the
implementation of Mueller's idea in the scalar model.
In Sect.~\ref{residue} we obtain the exact expression for the double
on-mass-shell residue of the instanton propagator. In Sect.~\ref{nu0} 
we discuss the leading and propagator corrections to the exponential 
of the cross section in the limit $\nu \rightarrow 0$. An analytical 
expression for them is obtained. We also discuss the 
terms in $\nu$ for this concrete example. In Sect.~\ref{general} 
the propagator correction to the function $F(\epsilon,\nu)$ for a wide 
range of $\epsilon$ and $\nu$ is calculated numerically and analyzed 
in detail. In Sect.~\ref{approx} we discuss the consistency of 
the procedure used for the calculation of the propagator correction.
Sect.~\ref{conclusion} contains some discussion of the results and 
concluding remarks. 

\section{Perturbation theory around the instanton}
\label{PT}

Within the formalism of coherent states \cite{KRT} the cross section 
(\ref{sigma-N}) can be calculated using the formula 
\beq
\sigma_{N}(E) = \frac{1}{VTj} 
\sum_{a,b} \left| <b|S {\cal P}_{E} {\cal P}_{N} |a>
\right|^{2},             \label{sigma-N1}
\eeq
where $S$ is the $S$-matrix in the one-instanton sector and ${\cal P}_{E}$ 
and ${\cal P}_{N}$ are the projectors onto the subspaces of fixed energy 
$E$ and multiplicity $N$, respectively. The coherent states are 
characterized by complex variables $a_{\bk}$, $b_{\bk}$, where $\bk$ is 
the spatial momentum and the summation over coherent 
states in Eq. (\ref{sigma-N1}) stands for the functional integral 
\[
\sum_{a,b} \ldots \equiv \int Da_{\bk} Da_{\bk}^{*} Db_{\bk} Db_{\bk}^{*} 
\exp \left( - \int d\bk a_{\bk}^{*} a_{\bk} -  
\int d\bk b_{\bk}^{*} b_{\bk} \right) \ldots  . 
\]
Using results of Refs. \cite{KRT}, \cite{Ti2} for the matrix elements 
of the $S$-matrix and the projectors ${\cal P}_{E}$, ${\cal P}_{N}$ in the 
coherent state representation we find that 
\bea
& & \sigma_{N}(E) = \int Da_{\bk} Da_{\bk}^{*} Db_{\bk} Db_{\bk}^{*} 
d\rho d\rho' d^{4}x_{0} d^{4}\xi d\theta 
 \exp \left[ - a^{*} a - b^{*} b - iP\xi - iN\theta \right]
\nonumber \\ 
 & & \times \exp \left[
\Gamma_{\rho}\left(b_{\bk}^{*}e^{ikx_{0}},
a_{\bk}e^{-ikx_{0}+ik\xi+i\theta} \right) + 
\Gamma_{\rho'}^{*}\left(b_{\bk},a_{\bk}^{*}\right) \right], 
\label{sigma-N2}
\eea
where $kx_{0} = w_{\bk}(x_{0})_{0} - \bk {\bf x}_{0}$, 
$w_{\bk}=\sqrt{\bk^{2}+m^{2}}$. 
Among the parameters of the family of instanton solutions we explicitly 
indicate the position of the center of the instanton denoted here by 
$x_{0}$, whereas $\rho$ stands for the rest of the parameters. 
The parameter $x_{0}$ corresponds to the translational symmetry 
of the theory. The term $(-iP\xi)$, $P_{\mu} = (E,\bf{P})$ is the 
total 4-momentum of the initial state, comes from the matrix element 
of the projector ${\cal P}_{E}$. Correspondingly, the term 
$(-iN\theta)$ in the exponent appears due to the projector 
${\cal P}_{N}$. See Refs. \cite{Ti1}, \cite{KRT} for the details. 
The ``effective action'' $\Gamma_{\rho}(b_{\bk}^{*},a_{\bk})$ to the 
leading and next-to-leading orders has the following form: 
\bea
\Gamma_{\rho}(b_{\bk}^{*},a_{\bk}) & = & - S_{inst}(\rho)
+ \Gamma_{1}(b_{\bk}^{*},a_{\bk}) + \Gamma_{2}(b_{\bk}^{*},a_{\bk}) + \ldots, 
\label{Gamma} \\
\Gamma_{1}(b_{\bk}^{*},a_{\bk}) & = &  
 a R_{a} +  b^{*} R_{b} + ab^{*}, \label{Gamma-1} \\
\Gamma_{2}(b_{\bk}^{*},a_{\bk}) & = & 
\frac{1}{2} a R_{aa} a + a R_{ab} b^{*} + \frac{1}{2} b^{*} R_{bb} b^{*}.    
\label{Gamma-2} 
\eea
The dots in (\ref{Gamma}) stand for higher order terms in $a_{\bk}$, 
$b_{\bk}^{*}$. The integration over the spatial momentum is implicit, 
so that $ a R_{a} \equiv \int d \bk a_{\bk} R_{a}(\bk)$, etc. 

The functions $R_{a}(\bk)$ and $R_{b}(\bk)$ in Eq. (\ref{Gamma-1}) are 
proportional to the residue of the Fourier transform 
of the instanton field continued to the Minkowski values of 
$k_{0}$ and taken on the mass shell. Their definitions are 
the following: 
\bea
R_{a}(\bk) & = & \frac{1}{(2\pi)^{3/2}} \frac{1}{\sqrt{2 w_{\bk}}} 
(k^{2} + m^{2}) \tilde{\phi}_{inst}(k_{0},\bk;0,\rho) |_{k_{0}=iw_{\bk}}, 
\label{Ra-def} \\
R_{b}(\bk) & = & \frac{1}{(2\pi)^{3/2}} \frac{1}{\sqrt{2 w_{\bk}}} 
(k^{2} + m^{2}) \tilde{\phi}_{inst}(k_{0},-\bk;0,\rho) |_{k_{0}=-iw_{\bk}}.  
\label{Rb-def} 
\eea
The Fourier transform of the instanton field is defined 
in the standard way:
\beq
 \tilde{\phi}_{inst}(k;x_{0},\rho) = \int d^{4}x \phi_{inst}(x;x_{0},\rho) 
 e^{ikx},    \label{FTinst-def}
\eeq
where we explicitly indicated the dependence of the instanton solution 
on $x_{0}$ and $\rho$. The function $R_{a}(\bk)$ 
corresponds to the initial state, whereas $R_{b}(\bk)$ corresponds 
to the final state. Let us denote the residue of the Fourier transform 
of the instanton at $x_{0}$ as $I(\rho)$:
\beq
I(\rho) = (k^{2}+m^{2}) \tilde{\phi}_{inst}(k;0,\rho)|_{k^{2}=-m^{2}}.
\label{Iinst-def}
\eeq
It is clear that for a scalar theory $I(\rho)$ is independent of $\bk$
and 
\beq
R_{a}(\bk) = R_{b}(\bk) = \frac{1}{(2\pi)^{3/2}} \frac{1}{\sqrt{2w_{\bk}}}
I(\rho).  \label{RaRb-I}
\eeq

The functions 
$R_{aa}$, $R_{ab}$ and $R_{bb}$ in (\ref{Gamma-2}) are 
defined through the double on-mass-shell residue of the 
Fourier transform of the propagator in the instanton background: 
\bea
R_{aa}(\bk,\bq) & = & \frac{1}{(2\pi)^{3}} \frac{1}{2\sqrt{w_{\bk}w_{\bq}}} 
(k^{2}+m^{2}) (q^{2}+m^{2}) 
G(k_{0},\bk;q_{0},\bq;0,\rho)|_{\stackrel{k_{0}=iw_{\bk}}{q_{0}=iw_{\bq}}},
\nonumber  \\
R_{ab}(\bk,\bq) & = & \frac{1}{(2\pi)^{3}} \frac{1}{2\sqrt{w_{\bk}w_{\bq}}} 
(k^{2}+m^{2}) (q^{2}+m^{2}) 
G(k_{0},\bk;q_{0},-\bq;0,\rho)|_{\stackrel{k_{0}=iw_{\bk}}{q_{0}=-iw_{\bq}}},
\label{Rab-def} \\
R_{bb}(\bk,\bq) & = & \frac{1}{(2\pi)^{3}} \frac{1}{2\sqrt{w_{\bk}w_{\bq}}} 
(k^{2}+m^{2}) (q^{2}+m^{2}) 
G(k_{0},-\bk;q_{0},-\bq;0,\rho)|_{\stackrel{k_{0}=-iw_{\bk}}
{q_{0}=-iw_{\bq}}}.
\nonumber
\eea
The instanton propagator $G(x,y;x_{0},\rho)$ is 
determined by the operator $\hat{D}_{x}$ of quadratic fluctuations 
around the instanton (see Eq. (\ref{GF-2})). Calculation 
of the propagator and of its residue in the scalar model will be 
the main subject of discussion in Sects.~\ref{model}-\ref{residue}. 
The Fourier transform of $G(x,y;x_{0},\rho)$  
is defined in the standard way: 
\beq
G(k,q;x_{0},\rho) = \int d^{4}x d^{4}y e^{ikx+iqy} G(x,y;x_{0},\rho). 
\label{FTprop-def}
\eeq
Again, $R_{aa}$ corresponds to the propagator connecting particles 
of the initial state, $R_{ab}$ connects a particle of the initial state 
with a particle of the final states and $R_{bb}$ connects two 
particles of the final state. It is easy to see that the dependence of 
the double on-mass-shell residue of the propagator on the momenta $\bk$, 
$\bq$ is of the form $(\bk,\bq)$. In a scalar model the momentum 
dependence of the Fourier transform of the instanton propagator 
is simple. For example, one can choose $k^{2}$, $q^{2}$ and $s=(k+q)^{2}$ 
as variables. Then the double on-mass-shell residue of the 
propagator for all three formulas (\ref{Rab-def}) can be expressed in terms 
of a single function $\tilde{P}(s;\rho)$, and one gets 
\beq
R_{\#}(\bk,\bq) = \frac{1}{(2\pi)^{3}} \frac{1}{2\sqrt{w_{\bk}w_{\bq}}} 
\tilde{P} \left(s_{\#}(\bk,\bq);\rho \right), 
\label{RRgen}
\eeq
where $\# = aa$, $ab$, $bb$, the $s$-variables are defined by 
\bea
s_{aa}(\bk, \bq) & = & s_{bb}(\bk, \bq) =
- 2 m^{2} - 2(\omega_{\bk} \omega_{\bq} - \bk\bq ), \label{saa} \\
s_{ab}(\bk, \bq) & = & 
- 2 m^{2} + 2(\omega_{\bk} \omega_{\bq} - \bk\bq ). \label{sab}
\eea  
in accordance with definitions (\ref{Rab-def}), and the function 
$\tilde{P}$ is given, for example, by 
\beq
\tilde{P} \left( s_{aa}(\bk,\bq);\rho \right) = (k^{2}+m^{2}) (q^{2}+m^{2})
G(k,q;0,\rho)|_{\stackrel{k_{0}=iw_{\bk}}{q_{0}=iw_{\bq}}}.
\label{Pinst-def}
\eeq

The integral over the variables $a_{\bk}$, $a_{\bk}^{*}$,  
$b_{\bk}$ and $b_{\bk}^{*}$ in the leading and next-to-leading 
approxmations is Gaussian and can be readily performed. Indeed, we can 
write the part of (\ref{sigma-N2}), including these 
variables, as 
\beq
 Z = \int D \psi_{\bk} 
\exp \left[- \frac{1}{2} \psi^{T} M \psi + J^{T}\psi \right], 
\label{psi-int}
\eeq
where $\psi_{\bk}^{T} = (a_{\bk},a_{\bk}^{*},b_{\bk},b_{\bk}^{*})$, and the 
matrix $M$ and the source $J$ can be read from (\ref{sigma-N2}), 
(\ref{Gamma}) - (\ref{Gamma-2}). In particular, 
\beq
M = M^{(1)} + M^{(2)} + \ldots ,   \label{M12}
\eeq
where $M^{(1)}$, corresponding to the leading contribution, includes the 
terms that do not contain the propagator residue, whereas $M^{(2)}$ 
contains propagator contributions of Eq. (\ref{Gamma-2}). 
As we will see later, in the scalar model 
$M^{(1)} \sim (m\rho)^{2}$, where $\rho$ is the size of the instanton. 
For the range of energies that will be studied in Sect.~\ref{nu0} and 
Sect.~\ref{general} $(m\rho)$ is a small parameter. Then  
$M^{(2)} \sim (m\rho)^{4}$ is indeed a small correction comparing to 
the leading term. In the leading approximation integration over $\psi$ 
gives 
\[
 Z \sim \exp \left[ \frac{1}{2} \tilde{\psi}^{T} M^{(1)} \tilde{\psi} \right]
= \exp \left[ \frac{1}{2} J^{T} \left( M^{(1)}\right)^{-1} J \right], 
\]
where $\tilde{\psi}$ is the saddle point value 
\beq
\tilde{\psi} = \left( M^{(1)}\right)^{-1} J,   \label{psi-saddle}
\eeq
calculated in the leading order approximation. Then 
the cross section in the leading approximation is equal to \cite{Ti2}
\beq
\sigma_{N}(E) =  \int d\rho d\rho' d^{4}x_{0} d^{4}\xi d\theta 
 \exp \left[ - S_{inst}(\rho) - S_{inst}(\rho') + 
\frac{1}{\lambda} W^{(1)}(x_{0},\rho,\rho',\xi,\theta) 
\right],  \label{sigma-N3}
\eeq
where 
\bea
 & & \frac{1}{\lambda} W^{(1)}(x_{0},\rho,\rho',\xi,\theta) =
 - i P \xi - i N \theta + \frac{1}{2} J^{T} \left( M^{(1)}\right)^{-1} J 
  \label{W1}\\
 & & = - i P \xi - i N \theta + R_{b}^{*} \frac{T}{1 - \gamma X} R_{b}
  + \gamma R_{a}^{*} \frac{X T^{-1}}{1 - \gamma X} R_{a} \nonumber \\
  & & + \gamma R_{a} \frac{X}{1 - \gamma X} R_{b} +
 \gamma R_{a}^{*} \frac{X}{1 - \gamma X} R_{b}^{*}. \nonumber
\eea
Here we introduced the variable 
\[
\gamma = \exp (i\theta)
\]
and the functions
\beq
X(\bk,\bq) = \delta (\bk - \bq) e^{iw_{\bk}\xi}, \; \; \; 
T(\bk,\bq) = \delta (\bk - \bq) e^{iw_{\bk}x_{0}},   \label{XT-def}
\eeq
which are treated as matrices in the momentum space, so that the 
integrations in Eq. (\ref{W1}) are viewed as matrix multiplications.  

In the next-to-leading approximation the cross section is given by 
\bea
\sigma_{N}(E) & = & \int d\rho d\rho' d^{4}x_{0} d^{4}\xi d\theta 
 \exp \left[ - S_{inst}(\rho) - S_{inst}(\rho') + 
\frac{1}{\lambda} W^{(1)}(x_{0},\rho,\rho',\xi,\theta)
  \right. \nonumber \\
 & + & \left. \frac{1}{\lambda} W^{(2)}(x_{0},\rho,\rho',\xi,\theta) 
\right],  \label{sigmaN-2}
\eea
where the contribution $W^{(2)}$ is obtained by evaluation 
of the term 
\[
 - \frac{1}{2} \psi^{T} M^{(2)} \psi 
\]
(see (\ref{psi-int}), (\ref{M12})) at $\psi = \tilde{\psi}$ given by 
the leading order value (\ref{psi-saddle}). This is sufficient 
for the accuracy with which the next-to-leading order is calculated. 
One can check that if more exact expression 
$\tilde{\psi} = (M^{(1)} + M^{(2)})^{-1} J$ is used, then correction 
terms contribute already to the next-next-to-leading
order. The next-to-leading order contribution 
\beq
\frac{1}{\lambda} W^{(2)} = - \frac{1}{2} \tilde{\psi}^{T} M^{(2)} 
 \tilde{\psi}   \label{W2-0}
\eeq
can be written as the sum of partial contributions involving 
the propagator between final states, between initial and final 
states and between initial states, respectively:
\beq
W^{(2)} =  W^{(2)}_{(f-f)} + W^{(2)}_{(i-f)} + W^{(2)}_{(i-i)}.
\label{W2}
\eeq 
Calculating (\ref{W2-0}) one obtains the following expressions 
for the partial contributions \cite{Ti2}:
\bea
\frac{1}{\lambda} W^{(2)}_{(f-f)} & = &  \frac{1}{2}
 R_{b} \frac{T}{1-\gamma X} R_{bb}^{\dagger}\frac{T}{1-\gamma X}R_{b}
                     \label{W2-ff}  \\
& + & \gamma R_{b}^{*} \frac{T}{1-\gamma X} R_{bb}\frac{X}{1-\gamma X} R_{a}
 + \frac{\gamma^{2}}{2}
 R_{a} \frac{X}{1-\gamma X} R_{bb} \frac{X}{1-\gamma X}R_{a}
 + \mbox{h.c.},  \nonumber \\
\frac{1}{\lambda} W^{(2)}_{(i-f)}& = & \gamma
R_{b} \frac{T}{1-\gamma X} R_{ab}^{\dagger}
\frac{X T^{-1}}{1-\gamma X} R_{a}
        \label{W2-if}  \\
& + & \gamma R_{b} \frac{X}{1-\gamma X} R_{ab} \frac{T}{1-\gamma X} R_{b}^{*}
 + \gamma^{2} R_{a}^{*} \frac{X T^{-1}}{1 - \gamma X} D_{ab}
\frac{X}{1-\gamma X} R_{a}  \nonumber \\
& + & \gamma^{2} R_{b} \frac{X}{1 - \gamma X} R_{ab} \frac{X}{1-\gamma X}
R_{a} + \mbox{h.c.},     \nonumber \\
\frac{1}{\lambda} W^{(2)}_{(i-i)}& = & \frac{\gamma^{2}}{2}
 R_{a} \frac{X T^{-1}}{1-\gamma X} R_{aa}^{\dagger}
\frac{X T^{-1}}{1-\gamma X}R_{a}
             \label{W2-ii} \\
& + & \gamma^{2} R_{a} \frac{X T^{-1}}{1-\gamma X} R_{aa}^{\dagger}
\frac{X}{1-\gamma X} R_{b}^{*}
 +  \frac{\gamma^{2}}{2} R_{b} \frac{X}{1 - \gamma X} R_{aa}
\frac{X}{1-\gamma X} R_{b}  + \mbox{h.c.}. \nonumber
\eea 

The next step is to evaluate the integrals in (\ref{sigmaN-2}). 
This can be done by the saddle point method.
Again, one can check that for the calculation of the cross section 
to the leading and next-to-leading orders it is enough to use the saddle 
point values of the parameters determined by the leading-order saddle point
equations. These equations are obtained by differentiation of the expression
\beq
 -S_{inst}(\rho) - S_{inst}(\rho') + 
\frac{1}{\lambda} W^{(1)}(x_{0},\rho,\rho',\xi,\theta),  \label{saddle-W1}
\eeq
where $W^{(1)}$ is given by Eq. (\ref{W1}) 
with respect to $x_{0}$, $\rho$, $\rho'$, $\xi$ and $\theta$. 
Physically relevant saddle points have ${\bf x}_{0}=0$, 
${\bf \xi}=0$ and $\rho = \rho'$, while $(x_{0})_{0}$, $\xi_{0}$ and
$\theta$ are purely imaginary. Therefore, it is convenient to introduce the
following notations:
\beq
(x_{0})_{0}= i\tau, \; \; \; \xi_{0} = i\chi, \; \; \;
\theta = -i \ln \gamma.   \label{x-tau}
\eeq
We will denote the saddle point values of the parameters 
$\tau m$, $\chi m$,$\rho$ and $\gamma$ as 
$\tilde{\tau}$, $\tilde{\chi}$, $\tilde{\rho}$ and $\tilde{\gamma}$ 
respectively. 
One can easily see that they are functions of $\epsilon = E/E_{sph}$ and 
$\nu = N/N_{sph}$. 

In the subsequent sections we will derive the saddle point equations
explicitly, find their solutions and calculate the leading and 
next-to-leading corrections to the cross section in the 
scalar model. Here we would like to study some general features 
of these corrections.

Suppose that the asymptotic behaviour of 
$R_{a}(\bk)$ and $R_{b}(\bk)$ for large $|\bk|$ is of the form 
\[
R_{a,b}(\bk) \sim |\bk|^{\delta}
\]
with $\delta \geq -1/2$. This assumption is verified in concrete examples. 
For example, in the scalar theory $\delta = -1/2$. As it has
been already explained in the Introduction,  we will be particularly
interested in the limit $\nu \rightarrow 0$. It can be shown 
(see Ref. \cite{Ti2}) that in this limit  
\beq
\tilde{\chi} - \tilde{\tau} \sim \nu \; \; \; \mbox{and} 
\; \; \; \tilde{\gamma} \sim \nu^{2\delta + 4}.      \label{chi-tau}
\eeq 
Estimating the integrals in expression (\ref{saddle-W1}), one can see that 
the terms that do not vanish in the limit $\nu \rightarrow 0$ 
can be written as 
\beq
\hat{W} = \tilde{\epsilon} \chi m  - \tilde{\nu} \ln \gamma 
- 2 \lambda S_{inst} (\rho) + \lambda \left[ \gamma R_{a} XT^{-1} R_{a} 
+ R_{b} T R_{b} \right],    \label{saddle-W1-1}
\eeq
where we found convenient to introduce the variables 
\beq
\tilde{\epsilon} = \frac{\lambda}{m} E, \; \; \; \tilde{\nu} = N \lambda.  
\label{teps-tnu-def}
\eeq
They are related to $\epsilon$ and $\nu$ by obvious relations 
\beq
\epsilon = \frac{1}{\kappa}\tilde{\epsilon}, \; \; \; 
\nu = \frac{1}{\kappa'}\tilde{\nu}  \label{eps-nu-rel}
\eeq
(see Eqs. (\ref{E-sph}), (\ref{N-sph})). 
The saddle point equations are 
\bea
\frac{\partial \hat{W}}{\partial \rho} & = & 
- 2 \lambda \frac{\partial}{\partial \rho} S_{inst}(\rho) + 
  \lambda  \frac{\partial}{\partial \rho}  
\left[ \gamma R_{a} XT^{-1} R_{a} + R_{b} T R_{b} \right],    
      \label{saddle-zeta-gen}  \\
\frac{\partial \hat{W}}{\partial \gamma} & = & -\frac{\tnu}{\gamma} 
+ \lambda R_{a} XT^{-1}R_{a} = 0,   \label{saddle-g-gen}  \\
\frac{\partial \hat{W}}{\partial \chi} & = & \teps  
 - \lambda \gamma R_{a} |\bk| XT^{-1}R_{a} = 0,   \label{saddle-chi-gen}  \\ 
\frac{\partial \hat{W}}{\partial \tau} & = & \lambda \left[ 
\gamma R_{a} |\bk| XT^{-1}R_{a} - R_{b} w_{\bk} T R_{b} \right]= 0. 
\label{saddle-tau-gen}   
\eea 
Here we used the definitions (\ref{XT-def}), (\ref{x-tau}). 
Note that the products in Eqs. (\ref{saddle-chi-gen}) and  
(\ref{saddle-tau-gen}) contain insertions $|\bk|$ and $w_{\bk}$. They are 
understood as additional factors in the integrand. For example, 
the last term in (\ref{saddle-tau-gen}) stands for  
\[
\int d\bk R_{b}(\bk) w_{\bk} e^{-w_{\bk}\tau} R_{b}(\bk).  
\]
In the limit $(\tilde{\chi} - \tilde{\tau}) \sim \tnu \rightarrow 0$ 
the main contribution in the integral 
\[
\tilde{\gamma} R_{a} XT^{-1}R_{a} = \tilde{\gamma} \int d\bk R_{a}(\bk)
e^{-w_{\bk}(\tilde{\chi}-\tilde{\tau})} R_{a}(\bk) 
\]
comes from large $|\bk| \sim (\tilde{\chi}-\tilde{\tau})^{-1}$, 
and up to negligible corrections 
$w_{\bk}$ can be substituted by $|\bk|$, i.e. the mass can be neglected. 
Analyzing Eqs. (\ref{saddle-g-gen}) and (\ref{saddle-chi-gen}) it 
is easy to see that properties (\ref{chi-tau}) indeed take place. 

In the next-to-leading order the terms which give non-vanishing 
contributions in the limit $\nu \rightarrow 0$ are the following: 
\beq
W^{(2)} = \lambda \left[ \gamma^{2} R_{a} XT^{-1} R_{aa} XT^{-1} R_{a} + 
2\gamma R_{a} XT^{-1} R_{ab} T R_{b} + R_{b} T R_{bb} T R_{b} \right] + 
{\cal O}(\nu).   \label{saddle-W2}
\eeq
Evaluating expressions for $\hat{W}$ and $W^{(2)}$, Eqs. 
(\ref{saddle-W1-1}) and (\ref{saddle-W2}), at the saddle point solutions 
of Eqs. (\ref{saddle-zeta-gen}) - (\ref{saddle-tau-gen}) we obtain 
the functions $F^{(1)}(\epsilon, \nu)$ and $F^{(2)}(\epsilon, \nu)$, 
respectively, in the limit $\nu \rightarrow 0$. These are precisely 
the leading and next-to-leading (propagator) corrections to the function 
$F(\epsilon,\nu)$.

The integrals in Eq. (\ref{saddle-W2}) can be represented in a general 
form as   
\beq
R_{\#1} O_{1} R_{\#1 \#2} O_{2} R_{\#2} = \int d\bk d\bq R_{\#1}(k) 
O_{1}(k) R_{\#1 \#2}(k,q,\theta) O_{2}(q) R_{\#2}(q),   \label{int-gen}
\eeq
where $\#1,\#2=a,b$ (so that $\#1 \#2 = aa$ or $ab$, or $bb$) and
the variable $\theta$ is the angle between 
$\bk$ and $\bq$. This integral transforms to 
\bea
& & 128\pi^{4} \int_{0}^{\infty} |\bk|^{2}d|\bk| 
\int_{0}^{\infty} |\bq|^{2}d|\bq| 
R_{\#1}(|\bk|) O_{1}(|\bk|) \nonumber \\
& & \times \left( \int_{0}^{\pi} \sin \theta d\theta 
\frac{R_{\#1 \#2}(\bk,\bq)}{16 \pi^{2}} \right) O_{2}(|\bq|) R_{\#2}(|\bq|) 
\label{int-gen1}  \\
& & = \frac{16\pi^{2}}{I^{2}(\rho)} \int d\bk d\bq R_{\#1}(|\bk|) 
O_{1}(|\bk|) R_{\#1}(|\bk|) {\cal S}_{\#1 \#2}(|\bk|,|\bq|) R_{\#2}(|\bq|) 
O_{2}(|\bq|) R_{\#2}(|\bq|),
\nonumber 
\eea
where the function ${\cal S}_{\#}(|\bk|,q)$ was defined as 
\beq
{\cal S}_{\#}(|\bk|,|\bq|) = \int_{0}^{\pi} \sin \theta d\theta 
\frac{R_{\#}(\bk,\bq))}{16\pi^{2}}  \label{cS-def}
\eeq   
for $\# = aa,ab,bb$. 

Singular contributions to the propagator correction at $\nu \rightarrow 0$ 
come from the first two terms in the r.h.s. of Eq. (\ref{saddle-W2}). 
Indeed, as one can see from a simple analysis, due to the presence 
of the $XT^{-1}$ factors either one of the momenta of integration or both 
are $\sim m(\tchi - \ttau)^{-1} \sim m/\nu$ and are large. Hence, the 
asymptotic form of ${\cal S}_{\#}(k,q)$ can be used. 

The leading large momentum asymptotics of the Fourier transform of the 
instanton propagator was calculated in Ref. \cite{Vo1} and is equal to 
\beq
G(k,q;\rho) = n'(\rho) \tilde{\psi}_{\mu}(k) \tilde{\psi}_{\mu}(q) 
\ln \frac{s}{m^{2}} + \ldots ,  \label{Gas-0}
\eeq
the factor $n'(\rho)$ being independent of the momenta.   
$\tilde{\psi}_{\mu}(k)$ is the 
Fourier transform of the zero translational mode. In a scalar theory 
this function can be written as $\tilde{\psi}_{\mu}(k) = 
k_{\mu} h(k^{2})/(k^{2}+m^{2})$. Then 
\[
\tilde{P}(s_{\#}(\bk,\bq);\rho) = n(\rho) s_{\#}(\bk,\bq) 
\ln \frac{s_{\#}(\bk,\bq)}{m^{2}} + \ldots 
\]
with $n(\rho)=n'(\rho)h^{2}(-m^{2})$. From this it follows that when 
at least one of arguments is large the leading asymptotics of 
${\cal S}_{\#}(|\bk|,|\bq|)$ can 
be written as a sum of factorized terms. We obtain: 

when $k,q \rightarrow \infty$
\beq
{\cal S}_{\#}(|\bk|,|\bq|) =  \frac{n(\rho)}{16\pi^{2}} 4 l_{\#} |\bk||\bq|  
\left[ \ln \frac{|\bk|}{m} + \ln \frac{|\bq|}{m} + \ldots \right], 
\label{cS-as1} 
\eeq

when $q \rightarrow \infty$ and $k$ is finite
\beq
{\cal S}_{\#}(|\bk|,|\bq|) = \frac{n}{16\pi^{2}} 4 l_{\#} w_{\bk} |\bq| 
\left[\ln \frac{|\bq|}{m} + \ldots \right].  \label{cS-as2} 
\eeq
Here $l_{aa}=l_{bb}=-1$ and $l_{ab}=1$.

Now let us analyze the first term in the r.h.s. of Eq. (\ref{saddle-W2}), 
which corresponds to the contribution of initial states. It gives 
\bea
& & F^{(2)}_{(i-i)}(\epsilon,\nu) = 
\lambda \tgamma^{2} R_{a} XT^{-1} R_{aa} XT^{-1} R_{a} \label{W2-aa-gen} \\
& & =  \lambda \frac{64 \pi^{2}}{I^{2}(\rho)} \tgamma^{2} 
\left(R_{a}XT^{-1} \frac{|\bk|}{m} \ln \left(\frac{|\bk|}{m} \right) 
R_{a} \right)_{\bk} \cdot 
\left(R_{a} XT^{-1} \frac{|\bq|}{m} R_{a} \right)_{\bq} + \ldots,
\nonumber    
\eea
where asymptotical formula (\ref{cS-as1}) was used. 

Using Eq. (\ref{cS-as2}) one can show that the second term in the r.h.s. of 
Eq. (\ref{saddle-W2}), which is due to the initial-final states, 
gives the following partial contribution to the propagator correction: 
\bea
& & F^{(2)}_{(i-f)}(\epsilon,\nu) = 2\lambda \tgamma R_{a} XT^{-1} 
R_{ab} T R_{b} \label{W2-ab-gen} \\
& & = - \lambda \frac{64 \pi^{2}}{I^{2}(\rho)} \tgamma  
\left(R_{a}XT^{-1} \frac{|\bk|}{m} \ln \left(\frac{|\bk|}{m} \right) 
R_{a} \right)_{\bk} \cdot 
\left(R_{b} T \frac{w_{\bq}}{m} R_{b} \right)_{\bq} + \ldots. \nonumber     
\eea 
The subscript $\bk$ reminds that all factors of the corresponding 
integrand are taken at the momentum $\bk$ and are integrated over 
$\bk$. 
Singularities at $\nu \rightarrow 0$ are produced by the logarithmic terms:  
\[
\ln \frac{|\bk|}{m} \sim \frac{1}{\tchi -\ttau} \sim  \ln \frac{1}{\nu}. 
\]
Adding terms (\ref{W2-aa-gen}) and (\ref{W2-ab-gen}) together 
we obtain that 
\bea
& & F^{(2)}_{(i-f)} + F^{(2)}_{(i-f)} = 
 \lambda \frac{64 \pi^{2}}{I^{2}(\rho)}  
\left( \tgamma R_{a} XT^{-1} 
\frac{|\bk|}{m} \ln \left(\frac{|\bk|}{m} \right) R_{a} \right)_{\bk}    
\nonumber \\
& & \times \left[ \left( \tgamma R_{a} XT^{-1} 
\frac{|\bq|}{m}R_{a} \right)_{\bq} - 
\left(R_{b} T \frac{w_{\bq}}{m} R_{b}\right)_{\bq} \right] + \ldots   
\label{W2-aaab-gen} 
\eea  
In Eqs. (\ref{W2-aa-gen}), (\ref{W2-ab-gen}) and (\ref{W2-aaab-gen}) 
the dots stand for terms which are finite in the limit 
$\nu \rightarrow 0$. Due to the saddle equation (\ref{saddle-tau-gen}) 
the expression in the square brakets in Eq. (\ref{W2-aaab-gen}) is zero. 

We would like to emphasize that our proof of the cancellation of singularities 
$\ln (1/\nu)$ in the propagator correction is rather general. For the proof 
we, essentially, used the general structure of the expressions for 
$\hat{W}$ and $W^{(2)}$, the leading order saddle point equations
and the factorization property of $F^{(1)}$ and $F^{(2)}$. The latter 
follows from general formula (\ref{Gas-0}) for the leading asymptotics 
of the instanton propagator.  
   
\section{The model and the instanton propagator}
\label{model}

We consider the model of one component real scalar field, defined
by the Minkowskian action
\beq
S = \int d^{4}x \left[ \frac{1}{2} \left( \partial_{\mu} \phi
\right)^{2} - \frac{1}{2} m^{2} \phi^{2} + \frac{\lambda}{4!}
\phi^{4} \right],    \label{action}
\eeq
where $\lambda > 0$. The potential of the model is unbounded from
below, hence the minimum $\phi = 0$ is metastable. Underbarrier
tunelling of the initial state from this vacuum to the
instability region and its return to the trivial vacuum is the
transition which gives rise to the shadow process we are going
to study here. 

Let us consider first the case $m=0$. There is a well known instanton
solution in the massless theory given by the formula
\cite{Fu,Lip}
\beq
\phi_{inst}(x; x_{0}, \rho) = \frac{4\sqrt{3}}{\sqrt{\lambda}}
\frac{\rho}{(x-x_{0})^{2} + \rho^{2}}.   \label{inst}
\eeq
The solution is characterized by five parameters: four coordinates
of the center of the instanton $x_{0 \mu}$ and its size $\rho$. Due to
conformal invariance of the massless theory the action of the
instanton does not depend on its size,
\[
S_{inst}^{(0)} \equiv S(\phi_{inst}) = \frac{16 \pi^{2}}{\lambda}.
\label{S-inst}
\]

In the case $m \neq 0$ the mass term breaks the conformal invariance.
Using standard scaling arguments it can be shown that there are
no regular solutions of the Euclidean equations of motion with finite
action. The decay of the vacuum $\phi=0$ is dominated by a
constrained instanton, a configuration which can be regarded
as an approximate solution of the equations of motion. It minimizes
the action under the constraint that the size of the configuration
is $\rho$. A formalism for  construction of such configurations
and evaluation of the functional  integral was developed in
\cite{Aff}.

When $\rho m \ll 1$ the constrained instanton configuration
$\phi_{c.i.}(x)$ behaves like the instanton solution (\ref{inst}) 
of the massless theory at
$x \ll \rho$ and as a solution of the free massive theory,
\[
\phi_{c.i.}(x) \approx \frac{2 \sqrt{6 \pi}}{\sqrt{\lambda}}
\frac{\rho m}{\sqrt{|x|m}} \frac{e^{-m |x|}}{|x|},
\]
for $x > m^{-1}$. The action of such configuration is
\bea
S_{inst}(\rho) & = & \frac{16 \pi^{2}}{\lambda} - \frac{24 \pi^{2}}{\lambda}
(\rho m)^{2} \left[ \ln \frac{\rho^{2} m^{2}}{4} + 2 C_{E} + 1 \right]
 \label{S-ci} \\
& + & {\cal O} \left( (\rho m)^{4}, (\rho m)^{4}\ln (\rho m)^{2}, 
(\rho m)^{4}\ln^{2} (\rho m)^{2}\right),  
\eea
where $C_{E} = 0.577216 \ldots$ is the Euler constant. For the class
of constraints mentioned above the terms given in (\ref{S-ci}) do
not depend on the explicit form of the constraint, whereas the
correction ${\cal O}((\rho m)^{4})$  does. In
our analysis we limit ourselves to the constraint independent order
of the approximation. Therefore, all contributions to the propagator 
correction which are of the order $(\rho m)^{4}$ times, may be, some 
logarithmic factors or smaller are beyond the accuracy of our 
approximation. 

For $m \neq 0$ the potential barrier separating the trivial vacuum
$\phi =0$ from the instability region is finite. Its height is
characterized by a sphaleron solution, a static $SO(3)$-symmetric
configuration satisfying the equation of motion. In Ref. \cite{KT}
it was found that the energy of the  sphaleron is
\beq
E_{sph} = \kappa \frac{m}{\lambda}, \; \; \; \kappa = 113.4
\label{Esph}
\eeq
and the characteristic number of particles contained in the
sphaleron is given by
\beq
N_{sph} = \frac{\kappa'}{\lambda}, \; \; \; \kappa'=63       
\label{Nsph}
\eeq
(compare to Eqs. (\ref{E-sph}) and (\ref{N-sph})). 

Now let us consider the propagator in the instanton background in
the massless theory. It is defined by the operator $\hat{D}_{x}$
of quadratic fluctuations in the expansion of the
action around the instanton solution (see Eq. (\ref{GF-2})).
For the massless theory this operator is equal to
\[
\hat{D}_{x} = - \frac{\partial^{2}}{\partial x_{\mu}^{2}} +
\frac{\lambda}{2} \phi_{inst}^{2}(x;0,\rho) =
- \frac{\partial^{2}}{\partial x_{\mu}^{2}} + \frac{24 \rho^{2}}
{(\rho^{2} + x^{2})^{2}}.
\]
It can be easily seen that it possesses five zero modes
$\psi_{A}(x)$ $(A=1,2,3,4,5)$ corresponding to the translational
invariance and the scale invariance of the massless theory.
The zero modes can be obtained by differentiation
of the instanton solution with respect to the parameters $x_{0}$
and $\rho$:
\[
\psi_{A} \sim \frac{\partial}{\partial \zeta_{A}}
\phi_{inst}(x; x_{0}, \rho)|_{x_{0}=0}, \; \; \;
\zeta_{\mu} = (x_{0})_{\mu}, \; \; \zeta_{5}=\rho.
\]
In accordance with the general theory the propagator in the instanton
background is the inverse of $\hat{D}_{x}$ on the subspace of
functions orthogonal to functions $f_{A}(x)$ which satisfy the only
condition that the matrix
\[
\Omega_{AB} = \int dx \psi_{A} (x) f_{B}(x)   
\]
is invertible \cite{LevYaf}. Let us denote such propagator by 
$G_{f}(x,y)$. It satisfies the equation
\beq
\hat{D}_{x} G_{f}(x,y) = \delta(x-y) - \sum_{A} f_{A}(x) \Omega^{-1}_{AB}
\psi_{B}(x)   \label{prop-eqn}
\eeq
and the orthogonality constraints
\beq
\int dx f_{A}(x) G_{f}(x,y) = 0 = \int dy G_{f}(x,y) f_{B}(y).
\label{G-constr}
\eeq
The lower index $f$ reminds that the instanton propagator 
satisfies the constraints defined by $f(x)$. 
The r.h.s. of Eq. (\ref{prop-eqn}) is the projector onto the subspace
orthogonal to the functions $f_{A}(x)$. Thus, 
there is an ambiguity in the definition of the propagator due to the
choice of the functions $f_{A}(x)$, but physical results, of course, do
not depend on it. In the next section, following the ideas
of Refs.~\cite{Mu92,MMcLY}, we will use the freedom of choosing
constraints (\ref{G-constr}) to eliminate the leading asymptotics of
the instanton propagator and simplify the analysis of the
initial-state corrections.

Let us consider first a particularly simple and natural choice of
the functions $f_{A}(x)$, namely
\[
f_{A}(x) = w(x) \psi_{A}(x),    
\]
where the weight function
\[
w(x) = \frac{4 \rho^{2}}{(\rho^{2} + x^{2})^{2}}.
\]
Zero modes $\psi_{A}(x)$, orthonormal
with respect to this weight function, are equal to
\beq
\psi_{\mu}(x) = \frac{\rho}{\pi} \sqrt{\frac{15}{2}}
\frac{2 x_{\mu} \rho}{(\rho^{2} + x^{2})^{2}},
\; \; \;
\psi_{5}(x) = \frac{\rho}{\pi} \sqrt{\frac{15}{2}}
\frac{x^{2} - \rho^{2}}{(\rho^{2} + x^{2})^{2}}.
\label{0-modes-1}
\eeq
In this case $\Omega_{AB} = \delta_{AB}$. Allowing some abuse of notation 
we denote the propagator orthogonal to the zero modes themselves by
$G_{\psi}(x,y)$. It satisfies the equation
\beq
\left(- \frac{\partial^{2}}{\partial x_{\mu}^{2}} +
\frac{24 \rho^{2}}
{(\rho^{2} + x^{2})^{2}} \right) G_{\psi}(x,y) =
\delta(x-y) - w(x) \sum_{A=1}^{5} \psi_{A}(x) \psi_{A}(x).
\label{prop-eqn-1}
\eeq
Making use of the fact that the symmetry group of the (massless)
theory is the large conformal group $SO(5)$ one can reformulate
the theory as a free theory on the four-dimensional sphere $S^{4}$
\cite{Ku,Lip}. The relation between points on
$R^{4}$ and corresponding points on $S^{4}$ is
given by the standard stereographic projection.
The instanton propagator $G_{\psi}(x,x')$ on $R^{4}$ is related to the
(free) propagator $K_{\psi}(\xi,\xi')$ on $S^{4}$ by the formula \cite{Ku}
\beq
G_{\psi}(x,x') = \frac{4 \rho_{0}^{2}}{(\rho^{2} + x^{2})
(\rho^{2} + x^{'2})} K_{\psi}(\xi,\xi'),    \label{G-K}
\eeq
where the points $\xi$ and $\xi'$ on the sphere correspond to
the points $x$ and $x'$ in the Euclidean space respectively. In fact, 
it can be shown that the propagator $K_{\psi}$ depends only on the geodesic
distance $s(\xi,\xi')$ between the points. It is convenient
to introduce the function
\[
d(\xi,\xi') = \frac{1}{2} \left( 1 - \cos s(\xi, \xi') \right).
\]
As a function of the coordinates of
the  Euclidean space it is equal to
\beq
d(x,x') = \frac{\rho^{2}(x-x')^{2}}{(\rho^{2} + x^{2})
(\rho^{2} + x^{'2})}.    \label{d-def}
\eeq
The propagator $K_{\psi}(d)$ satisfies the equation
\beq
\left( - \Delta_{\xi} - 4 \right) K_{\psi}(d(\xi,\xi')) =
\delta(\xi,\xi') -
\sum_{A=1}^{5} \tilde{\psi}_{A}(\xi) \tilde{\psi}_{A}(\xi'),
\label{K-eqn}
\eeq
where $\Delta_{\xi}$ is the Laplacian, $\delta(\xi,\xi')$ is
the $\delta$-function on $S^{4}$, and the orthornormalized
zero-modes $\tilde{\psi}_{A}(\xi)$ of
the operator in the l.h.s. of Eq. (\ref{K-eqn})
are related to the zero modes $\psi_{A}(x)$, Eq. (\ref{0-modes-1}),
by
\[
\psi_{A}(x) = \frac{2 \rho}{(\rho^{2} + x^{2})}
\tilde{\psi}(\xi).
\]
If the sphere $S^{4}$ is considered as  being embedded into
the 5-dimensional Euclidean space with  coordinates $\{z_{A}\}$
$(A=1,2,3,4,5)$ then
$\tilde{\psi}_{A} = \sqrt{15/8 \pi^{2}} z_{A}$.

There are a few ways to find the instanton propagator $K_{\psi}(d)$ 
satisfying Eq.~(\ref{K-eqn}). Perhaps the simplest one is to start with
the expansion
\[
K_{\psi}(d(\xi,\xi')) = \sum_{l=0}^{\infty} \sum_{\{m\}}
\frac{1 - \delta_{l,1}}{\lambda_{l}^{2} - 4} Y_{l,m}(\xi)
Y_{l,m}^{*}(\xi').    
\]
The spherical harmonics $Y_{l,m}(\xi)$ on $S^{4}$ are
eigenfunctions of the Laplace operator with the eigenvalues
$\lambda_{l}$:
\bea
 - \Delta_{\xi} Y_{l,m}(\xi) & = & \lambda_{l} Y_{l,m}(\xi),
\nonumber \\
  \lambda_{l} & = & l(l+3), \; \; \; l=0,1,2, \ldots,
\nonumber
\eea
where $l$ is the total momentum and $m$ denotes the set of
orbital momentum numbers. Using summation formulas the explicit
expression for the instanton propagator on $S^{4}$
was obtained \cite{Ku}:
\[
K_{\psi}(d) = \frac{1}{8\pi^{2}} \left[ \frac{1}{2d} -
3 \ln d - \frac{43}{5} + 6 d \ln d + \frac{56}{5} d \right].
\]
Using Eq.~(\ref{G-K}) we arrive at the final
expression for the instanton propagator:
\bea
G_{\psi}(x,y) & = & \frac{1}{2\pi^{2}} \frac{\rho^{2}}
{(\rho^{2} + x^{2}) (\rho^{2} + y^{2})}
\left\{ \frac{1}{2d(x,y)} -
 \right.   \nonumber \\
& - & \left. 3 \ln d(x,y) - \frac{43}{5} + 6 d(x,y) \ln d(x,y) +
\frac{56}{5} d(x,y)  \right\},  \label{propag}
\eea
where the function $d(x,y)$ is given by Eq. (\ref{d-def}).

When points $x$ and $x'$, or the corresponding points $\xi$ and $\xi'$,
approach each other the propagator $G_{\psi}(x,x')$ is singular, the
leading singularity being given by the first term in the curly
brackets in the l.h.s. of Eq. (\ref{propag}). This is precisely the
singularity of the free propagator of the massless scalar theory
in the four-dimensional Euclidean  space. Indeed, this propagator,
denoted here by $G_{0}(x-x')$, satisfies the equation
\[
- \frac{\partial^{2}}{\partial x_{\mu}^{2}} G_{0}(x-x') = \delta (x-x')
\]
and is equal to
\beq
G_{0}(x-x') = \frac{1}{4 \pi^{2}} \frac{1}{(x-x')^{2}}.  \label{G0}
\eeq
The corresponding propagator $K_{0}(\xi,\xi')$
on $S^{4}$, related to $G_{0}(x-x')$ by the formula
similar to (\ref{G-K}), satisfies the following equation on the
four-dimensional sphere:
\[
\left( - \Delta_{\xi} + 2 \right) K_{0}(\xi,\xi') = \delta(\xi,\xi')
\]
(compare with Eq. (\ref{K-eqn})) and
is equal to
\[
K_{0}\left( \xi,\xi'\right) = \frac{1}{8\pi^{2}}
\frac{1}{2d(\xi,\xi')}.
\]
This gives the first term in the curly brackets in Eq. (\ref{propag}). 

Let us discuss briefly the structure of the expression 
for the instanton propagator. 
A detailed analysis shows \cite{Ku} that $G_{\psi}(x,x')$ can be written 
as the following infinite sum:  
\bea
G_{\psi}(x,x') & = & G_{0}(x-x') + \sum_{n=1}^{\infty} G_{n}(x,x')   
\nonumber \\ 
& = & G_{0}(x-x') + \frac{\lambda}{2}
\int dy G_{0}(x-y) \phi_{inst}^{2}(y) G_{0}(y-x') + \ldots
\label{G-Gn}
\eea
This series has an interpretation in terms of Feynman diagrams. 
The term $G_{n}(x,x')$ is represented as the diagram consisting of a 
line, corresponding to the free scalar propagator (\ref{G0}), with $n$ 
insertions $\phi_{inst}^{2}$. The first term in Eq. (\ref{G-Gn}) is 
just the free propagator without insertions. As we have already 
discussed, it gives the leading singularity of the full propagator 
$G_{\psi}(x,x')$ when $x \rightarrow x'$. The integral term, written down 
in Eq. (\ref{G-Gn}), is the $G_{1}(x,x')$ term. Calculating it we get  
\[
G_{1}(x,x') = - \frac{3}{2\pi^{2}} \frac{\rho^{2}}{(\rho^{2} + x^{2})
(\rho^{2} + x^{'2}) - \rho^{2}(x-x')^{2}}
\ln \frac{\rho^{2}(x-x')^{2}}{(\rho^{2} + x^{2})
(\rho^{2} + x^{'2})}. 
\]
This term has the logarithmic singularity when $x \rightarrow x'$. 
This is precisely the subleading logarithmic singularity of the exact 
expression (\ref{propag}). 

We finish this section by presenting a relation 
between the propagator $G_{f}(x,y)$, satisfying a general
constraint (\ref{G-constr}), and the propagator $G_{\psi}(x,y)$.
They are related as follows:   
\bea
G_{f}(x,y) & = & G_{\psi}(x,y) -
\left( \int dz G_{\psi}(x,z) f_{A}(z) \right) \Omega_{AB}^{-1}
\psi_{B}(y) \nonumber \\
 & - & \psi_{A}(x) \left( \Omega_{AB}^{T} \right)^{-1}
\int dz f_{B}(z) G_{\psi}(z,y)  \nonumber \\
& + & \psi_{A}(x) \left( \Omega_{AB}^{T} \right)^{-1}
\left( \int dz dz' f_{B}(z) G_{\psi}(z,z') f_{C}(z') \right)
\Omega_{CD}^{-1}  \psi_{D}(y).    \label{G-G}
\eea
In the next section we will see that with the help of this formula 
and by an appropriate choice of the function $f(x)$ one can modify the 
asymptotics of the instanton propagator. 

\section{The high energy asymptotics of the instanton propagator}
\label{asympt}

The Fourier transform of the instanton propagator is defined
by Eq. (\ref{FTprop-def}). 
In principle, using the exact expression (\ref{propag}) the 
function $G_{\psi}(p,q)$ can be obtained by direct calculation.
We did not find the complete analytical expression for it, instead we
derived the asymptotic formula for the Fourier transform of the
instanton propagator in the regime when $p^{2}$, $q^{2}$ are
fixed and $s \equiv (p+q)^2 \rightarrow \infty$. The growing terms
of the asymptotics are given by
\beq
G_{\psi}(p,q) = \frac{16 \pi^{2}}{p^{2} q^{2}} \left[
s\rho^{2} \ln (s \rho^{2}) \Pi_{1}(p,q) +
(s \rho^{2}) \Pi_{2}(p,q) +
\ln (s \rho^{2}) \Pi_{3}(p,q) + \ldots \right],   \label{G-asymp}
\eeq
where
\bea
\Pi_{1}(p,q) & = &  \frac{3}{4}  {\cal S}_{1}(p \rho)
{\cal S}_{1}(q \rho), \nonumber \\
\Pi_{2}(p,q) & = & \frac{3}{2} \left( C_{E} - \frac{1}{15}
- \ln 2 \right) {\cal S}_{1}(p \rho)
{\cal S}_{1}(q \rho), \nonumber \\
\Pi_{3}(p,q) & = & \left\{ {\cal S}_{1}(p\rho)
\left[ \frac{9}{2} {\cal S}_{2} (q\rho) - \left( \frac{27}{4} +
\frac{3}{4} q^{2}\rho^{2} \right) {\cal S}_{1}(q\rho) \right] \right.
\nonumber \\
& + &  \left.
\left[ \frac{9}{2} {\cal S}_{2} (p\rho) - \left( \frac{27}{4} +
\frac{3}{4} p^{2}\rho^{2} \right) {\cal S}_{1}(p\rho) \right]
{\cal S}_{1}(q\rho) - \frac{3}{2}
{\cal S}_{2}(p\rho) {\cal S}_{2}(q\rho) \right\}.  \nonumber
\eea
Here ${\cal S}_{n}(z)$ is defined by ${\cal S}_{n}(z)= z^{n} K_{n}(z)$, 
where $K_{n}(z)$ is the modified Bessel function. Using the explicit
expressions for the normalized translational zero modes Eq.
(\ref{0-modes-1}), one can easily see that the product of their 
Fourier transforms 
$\psi_{\mu}(p) \psi_{\mu}(q) \sim \rho^{2} s$. Thus, the first 
two terms of the asymptotic expansion (\ref{G-asymp}) can 
be written as
\beq
G_{\psi}(p,q) = -\frac{1}{5\rho^{2}} \ln (\rho^{2} s)
\psi_{\mu}(p) \psi_{\mu}(q)
 - \frac{2}{5 \rho^{2}} \left( C_{E} - \frac{1}{15}-\ln 2 \right)
\psi_{\mu}(p) \psi_{\mu}(q) + \ldots \label{G-asymp1}
\eeq
The leading term of the asymptotics of the propagator
in the instanton background was calculated in Ref. \cite{Vo1} and 
is given by Eq. (\ref{Gas-0}). 
This result is in complete agreement with the first term
in Eq. (\ref{G-asymp1}).

In Ref. \cite{Mu92} Mueller proposed the idea to use the ambiguity
in the choice of the function $f_{A}(x)$ in order 
to cancel the two leading terms in the asymptotics of the
propagator $G_{f}(p,q)$. If this can be done, 
then the propagator contribution and loop 
contributions of the initial state corrections vanish. As a 
consequence, such corrections do not exponentiate, i.e., do not 
give contributions to the function $F(\epsilon,\nu)$. 
Moreover, in this case the initial-final state corrections 
can be described semiclassically. Namely,  
the effect of the initial-state lines can be taken into
account by substituting the instanton by a new field
configuration which is a particular solution
to the classical equation of motion
with an external source (see Ref. \cite{Mu92} for details).

In the rest of this section we discuss how the functions $f_{A}(x)$ 
that provide vanishing of the two leading terms in Eq. (\ref{G-asymp}) 
(or Eq. (\ref{G-asymp1})) can be chosen. For this we essentially repeat the 
arguments of Ref. \cite{Mu92}. The propagator  
constraint (\ref{G-constr}) for which the
vanishing takes place turns out to be not relativistically covariant.
Let $p_{1}$ and $p_{2}$ be the arguments of the Fourier
transform of the propagator. We choose a coordinate system
such that the components $p_{1 j}=p_{2 j}=0$ for $j=2,3$, whereas 
\[
   p_{1+} \rho = p_{2}\rho \gg 1
\]
and $p_{1}^{2}$ and $p_{2}^{2}$ are fixed. Here the $\pm$ components
of the momenta are defined by
\[
p_{j\pm} = \frac{(p_{j})_{0}\pm(p_{j})_{1}}{\sqrt{2}}.
\]
Then $(p_{1},p_{2})\rho^{2} \approx p_{1+} p_{2-} \rho^{2} \gg 1$.
Only the components $f_{\mu}(x)$, corresponding to translations,
play a role. The Fourier transforms $\tilde{f}_{\mu}(p)$ of the functions
$f_{\mu}(x)$, defining the required propagator constraint, can be 
chosen in the following way:
\[
\tilde{f}_{\mu} (p) = \delta (p_{+}-M) \delta (p_{-}+M)
\bar{f}_{\mu} (p_{2},p_{3}),
\]
where $M$ is an arbitrary parameter of the dimension of mass.
Substituting such functions into Eq. (\ref{G-G}) one finds after
some calculation that indeed the $s \ln s$- and $s$-terms of the 
asymptotics (\ref{G-asymp}) cancel and 
\[
G_{f}(p_{1},p_{2}) = \psi_{\mu}(p_{1}) {\cal G}_{\mu \nu}
\psi_{\nu} (p_{2}) + \ldots ,
\]
where the $4 \times 4$ constant
matrix ${\cal G}_{\mu \nu}$ is equal to
\[
{\cal G}_{\mu \nu} = -\frac{1}{5 \rho^{2}} \Omega_{\mu \sigma}^{-1}
 \frac{1}{(2\pi)^{8}} \int d^{4}q_{1} d^{4}q_{2}
\tilde{f}_{\sigma}(-q_{1}) \psi_{\rho}(q_{1})
\ln \frac{(q_{1},q_{2})}{M^{2}} \psi_{\rho}(q_{2}) \tilde{f}_{\tau}(-q_{2})
\left( \Omega^{T} \right)_{\tau \nu}^{-1}.
\]
Using the freedom of choosing the function $\bar{f}_{\mu}$
one can make the constant real symmetric matrix
${\cal G}_{\mu \nu}$ equal to zero. We would like to stress 
that the knowledge of the exact expression for the instanton propagator 
allows us to get the explicit formula for the matric ${\cal G}_{\mu \nu}$.
This is in contrast with the case of the electroweak theory, where only a 
general structure of the analogous matrix can be derived \cite{Mu92}. 

\section{Residue of the instanton propagator}
\label{residue}

As we have seen in Sect.~\ref{PT} 
for the perturbative calculations of the function
$F(\epsilon,\nu)$ the on-mass-shell residue of the instanton
solution is needed. Its definition is given by Eq. (\ref{Iinst-def}). 
In the massless case the Fourier transform of the instanton solution is 
equal to 
\beq
  \tilde{\phi}_{inst}(p;0,\rho) = \frac{R_{0}}{\sqrt{\lambda}}
  \frac{\rho^{2}}{|p|} K_{1}(\rho |p|), \label{inst-FT}
\eeq
where $R_{0}=16 \sqrt{3} \pi^{2}$,
$K_{1}(z)$ is the modified Bessel function, and
\beq
  I(\rho) = \frac{\rho}{\sqrt{\lambda}} R_{0}.  \label{R-inst}
\eeq
Using definitions (\ref{Ra-def}), (\ref{Rb-def}) we obtain that 
\beq
R_{a}(\bk) = R_{b}(\bk) = \frac{1}{(2\pi)^{3/2} \sqrt{2w_{\bk}}} 
I(\rho) = \frac{\rho}{\sqrt{\lambda}} 
\frac{16 \sqrt{3} \pi^{2}}{(2\pi)^{3/2} \sqrt{2w_{\bk}}}.
\label{Ra-Rb}
\eeq
In massive theory one should take the Fourier transform of the constrained 
instanton solution. Later we will show that, in fact, within the 
approximation 
considered here it is enough to take the residue (\ref{R-inst}) 
of the massless instanton. Corrections due to non-zero mass 
are of the order ${\cal O}(\rho^{4}m^{4})$, i.e. of the order 
of terms already neglected in (\ref{S-ci}). However, in Eqs. 
(\ref{Ra-Rb}) we should use $w_{\bk}=\sqrt{\bk^{2}+m^{2}}$, 
i.e. the expression for the energy of the massive theory. 

To calculate the next-to-leading correction
to the function $F(\epsilon, \nu)$ we need the expression
for the double on-mass-shell residue of the instanton propagator.
The propagator residue is defined by Eq. (\ref{Pinst-def}). In the massles 
theory we have 
\beq
R_{\#}^{(0)}(\bk,\bq) = \frac{1}{(2\pi)^{3}} 
\frac{\rho^{2}}{2\sqrt{w_{\bk}w_{\bq}}} 
 P \left(\rho^{2} s^{(0)}_{\#}(\bk,\bq) \right), 
\label{RR}
\eeq
where $\# = aa$, $ab$, $bb$ and the function $s^{(0)}_{\#}(\bk,\bq)$ is 
the $s$-variable for the corresponding particles on the mass shell:
\beq
s^{(0)}_{aa}(\bk, \bq) = s^{(0)}_{bb}(\bk, \bq) = 
- s^{(0)}_{ab}(\bk, \bq) = 
- 2(|\bk| |\bq| - \bk\bq ).    \label{s-0}
\eeq
The functions $P(\rho^{2}s)$ and $\tilde{P}(s;\rho)$ in 
Eq. (\ref{Pinst-def}) are related in the following way: $\tilde{P}(s;\rho) = 
\rho^{2}P(\rho^{2}s)$. 
However, in the calculation of the 
next-to-leading order corrections due to non-zero mass  
must be taken into account. It turns out that within the accuracy set by 
Eq. (\ref{S-ci}) it is enough to use functions $R_{\#}(\bk,\bq)$ 
defined by the relation 
\beq
R_{\#}(\bk,\bq) = \frac{1}{(2\pi)^{3}} 
\frac{\rho^{2}}{2\sqrt{w_{\bk}w_{\bq}}}
 P \left(\rho^{2} s_{\#}(\bk,\bq) \right), 
\label{RR1}
\eeq  
where, as in the case of Eq. (\ref{RR}), $P$ is the residue of 
the instanton propagator of the {\it massless} theory, whereas 
the energy and the $s$-variables are taken for the {\it massive} one. 
Namely, $w_{\bk}=\sqrt{\bk^{2}+m^{2}}$ and $s_{\#}(\bk,\bq)$ are given by 
Eqs. (\ref{saa}), (\ref{sab}). The motivation for such procedure 
of calculation is discussed in Sect.~\ref{approx}. 

Note, however, that there is an ambiguity in representation (\ref{RR1}). 
One can equally well substitute $s_{\#}^{(0)}(\bk,\bq)$ in (\ref{RR})
with any other function of $k$ and $q$ which coincides with  
$s_{\#}^{(0)}(\bk,\bq)$ for $k^{2}=q^{2}=0$. For example, one can use 
the expressions coming from $\tilde{s}=2(k,q)$, i.e. 
\bea
\tilde{s}_{aa}(\bk, \bq) & = & \tilde{s}_{bb}(\bk, \bq) =
- 2(\omega_{\bk} \omega_{\bq} - \bk\bq ), \label{tsaa} \\
\tilde{s}_{ab}(\bk, \bq) & = & 
2(\omega_{\bk} \omega_{\bq} - \bk\bq ). \label{tsab}
\eea
We will see later that the effect of this ambiguity is negligible for the 
final result.  
 
The {\it exact} expression for the function $P(\rho^{2} s)$ 
was obtained in Ref. \cite{KubTi}. 
Its calculation is a rather tedious although
straightforward procedure, and we do not give the details
here. Instead we would like to discuss some general features
of the computation before presenting the answer.

The terms in expression (\ref{propag}) give contributions
to the residue which can be divided in four classes.

1) The first term in the curly brackets in Eq. (\ref{propag}) gives
rise to the free propagator as it was already explained.  Its
contribution to the Fourier transform $G_{\psi}(k,q)$ is
proportional to
\[
\frac{2 (2 \pi)^{4}}{k^{2}} \delta (k + q).
\]
This describes free motion of the particle not interacting with the
instanton and is irrelevant for our problem.

2) There are factorizable terms of the form $g_{1}(x) g_{2}(y)$,
where $g_{i}(x)$'s are proportional to expressions like
\beq
 \frac{x^{n}}{(\rho^{2} + x^{2})^{l}} \; \; \; \mbox{or} \; \;
\; \frac{x^{n} \ln(\rho^{2} + x^{2})}{(\rho^{2} + x^{2})^{l}}
\label{fterms}
\eeq
with some integer $n$ and $l$. Their contributions to the
momentum-space propagator are of the form $\tilde{g}_{1}(k^{2})
\tilde{g}_{2}(q^{2})$. These are $s$-independent
contributions to the function $P(\rho^{2} s)$. 

3) The next group of terms is of the form $(xy) g_{1}(x) g_{2}(y)$
with $g_{i}$ of the form (\ref{fterms}). Calculating the momentum-space
propagator we get
\bea
 \int e^{ikx + iqy} (x,y) g_{1}(x) g_{2}(y) & = & -
\frac{\partial}{\partial k^{\mu}}
\frac{\partial}{\partial q^{\mu}}
\int e^{ikx + iqy} g_{1}(x) g_{2}(y) \nonumber \\
= - \frac{\partial}{\partial k^{\mu}}
\frac{\partial}{\partial q^{\mu}}
\tilde{g}_{1}(k^{2})
\tilde{g}_{2}(q^{2}) & = & - 4 (k,q)
\tilde{g}'_{1}(k^{2}) \tilde{g}'_{2}(q^{2}).
\nonumber
\eea
This gives a contribution to the residue proportional to
$s_{\#}(\bk,\bq)$.

4) The last group consists of terms of the form $(xy) \ln
(x-y)^{2} g_{1}(x) g_{2}(y)$ and $\ln (x-y)^{2} g_{1}(x)
g_{2}(y)$. Carrying out the computations one can show that they lead
to terms proportional to $s_{\#} \ln s_{\#}$ and $\ln s_{\#}$, 
respectively, in the expression for the residue of the instanton 
propagator. 

Finally, the {\it exact} expression for the function $P(z)$
is given by
\beq
 P(z) = 16 \pi^{2}  \left[ \frac{3}{4} z \ln \frac{z}{4} +
\frac{3}{2} z \left( C_{E} - \frac{1}{15}\right) -
\frac{3}{2}  \ln \frac{z}{4} - 
3 \left(C_{E} + \frac{43}{30}\right)
\right]. \label{R}
\eeq

Below this result will be used for the calculation
of the next-to-leading correction to the function
$F(\epsilon,\nu)$.
It is convenient to write Eq. (\ref{R}) as
\bea
P(z) & = & 16 \pi^{2} \left\{ \alpha_{1} \left[ z \ln \frac{z}{4} + 
2 \left(C_{E} - \frac{1}{15} \right) z \right] 
- \alpha_{2} \left[ \ln \frac{z}{4} + 
2 \left( C_{E} + \frac{43}{30} \right) \right] \right\},  \label{R1} \\
\alpha_{1} & = & 3/4, \; \; \; \alpha_{2}=3/2.   \nonumber 
\eea   
This form of the residue allows to trace the origin of various 
contributions to the final answer for the propagator correction. 

{}From Eq. (\ref{int-gen1}) we see that the important ingredient of the 
contributions to the propagator correction is the function 
${\cal S}_{\#}(|\bk|,|\bq|)$, defined by Eq. (\ref{cS-def}). 
With the residue $P(s)$ given by expression (\ref{R1})   
${\cal S}_{\#}(|\bk|,|\bq|)$ can be easily calculated. 
As it was already discussed in Sect.~\ref{PT}, when 
one of the arguments is large the asymptotics of 
this function can be represented as a sum of factorized terms. 
In the model under consideration we obtain  

when $k,q \rightarrow \infty$
\beq
{\cal S}_{\#}(|\bk|,|\bq|) \sim  4 l_{\#} |\bk||\bq| \rho^{2} \alpha_{1} 
\left[ \ln (|\bk|\rho) + \ln (|\bq|\rho) + 2\ln 2 - \frac{1}{2} + 
C_{A} \right], 
\label{cS-as1-model} 
\eeq

when $q \rightarrow \infty$ and $k$ is finite
\beq
{\cal S}_{\#}(|\bk|,|\bq|) \sim  4 l_{\#} w_{\bk} q \rho^{2} \alpha_{1} 
\left[\ln (|\bq|\rho) + \ln (|\bk|\rho) + \ln 2 + C_{A} + 
\frac{|\bk|}{w_{\bk}} M_{\#,2}(|\bk|,\infty) \right],  \label{cS-as2-model} 
\eeq
where $l_{aa}=l_{bb}=-1$, $l_{ab}=1$ and $M_{\#,2}(|\bk|,\infty)$ is 
given by Eqs. (\ref{Mgen2}), (\ref{Mab2}) in the Appendix. 
Let us recall that the leading terms in formulas (\ref{cS-as1-model}), 
(\ref{cS-as2-model}) were obtained in Sect.~\ref{PT} for a general model 
from the leading asymptotics of the instanton propagator 
of Ref. \cite{Vo1}. 
 
\section{Leading and propagator corrections: case $\nu \rightarrow 0$}
\label{nu0}

In this section, firstly, we study the saddle point solutions that 
give dominant contribution to integral (\ref{sigmaN-2}) in the limit 
$\nu \rightarrow 0$. Secondly, with this solution we calculate the 
leading and next-to-leading order contributions to $F(\epsilon,\nu)$. 
Finally, we give an illustration of cancellation of terms singular 
in $\nu$ in the case of the concrete model (\ref{action}).
 
For this model the leading contribution to $F(\epsilon,\nu)$ comes 
from the term  
\bea
\hat{W} & = & 48 \pi^{2} \rho^{2}m^{2} \ln(C \rho^{2}m^{2}) + 
\tilde{\epsilon} \chi m - \tilde{\nu} \ln \gamma         \label{W1-2} \\
& + & 192 \upsilon \rho^{2} \left[ J_{0}(\gamma,\tau,\chi) 
+ \gamma J_{0}(\gamma,\chi - \tau,\chi) + 
2\gamma J_{0}(\gamma,\chi,\chi) \right]   \nonumber 
\eea
(see Eqs. (\ref{sigma-N3}), (\ref{W1}) and (\ref{S-ci})). Here 
we introduced the notations 
\[
\ln C = -\ln 4 + 2C_{E} + 1 \approx 0.768137 \ldots, 
\; \; \; \upsilon = 192 \pi^{2}  
\] 
and 
\[
J_{0} (\gamma, \tau, \chi) = \frac{1}{4\pi} \int \frac{d \bk}{w_{\bk}} 
\frac{e^{-w_{\bk}\tau}}{1 - \gamma e^{-w_{\bk}\chi}}.  
\]
Recall that $\teps$ and $\tnu$ are defined by Eqs. (\ref{teps-tnu-def}) 
and are related to $\epsilon = E/E_{sph}$ and $\nu = N/N_{sph}$ 
by relations (\ref{eps-nu-rel}) with $\kappa$ and $\kappa'$ given by 
Eqs. (\ref{Esph}), (\ref{Nsph}).

In accordance with estimates (\ref{chi-tau}) we expect that in our 
model 
\beq
\tchi - \ttau \sim \nu, \; \; \mbox{and} \; \; 
\tgamma \sim \nu^{3}. \label{chi-tau-1}
\eeq
It will be shown shortly that this is ineed the case. 
In the leading approximation in $\nu$ integrals $J$  
in Eq. (\ref{W1-2}) can be taken at $\gamma=0$. Then they can be 
easily calculated, and one gets  
\beq
 \Phi (\tau m) \equiv \frac{1}{m^{2}} J(0,\tau,\chi) = 
 \frac{1}{\tau m} K_{1}(\tau m).     \label{Phi-def}
\eeq
Taking into account (\ref{chi-tau-1}) we find that in the leading 
order in $\nu$ the following expression for $\hat{W}$ can be used: 
\bea
\hat{W} & = & 48 \pi^{2} \rho^{2}m^{2} \ln (C \rho^{2}m^{2}) + 
\tilde{\epsilon} \chi m - 
 \tilde{\nu} \ln \gamma         \label{W1-3} \\
& + & \upsilon \rho^{2}m^{2} \left[ \Phi (\tau m) + 
\frac{\gamma}{m^{2}(\chi - \tau)^{2}} + {\cal O}(\nu^{3})  \right]. \nonumber 
\eea

Then the saddle point equations 
(\ref{saddle-zeta-gen}) - (\ref{saddle-tau-gen}) take the form 
\bea
\frac{\partial}{\partial \rho^{2}} \hat{W} & = & 
48 \pi^{2} m^{2}\ln \left( \rho^{2}m^{2} C e \right) + 
\upsilon m^{2} \left[ \Phi(\tau m) + 
\frac{\gamma}{m^{2}(\chi - \tau)^{2}}  \right], 
    \label{saddle0-r2} \\
\frac{\partial}{\partial \gamma} \hat{W} & = & 
- \frac{\tilde{\nu}}{\gamma} + \upsilon \rho^{2} 
\frac{1}{(\chi - \tau)^{2}}, 
    \label{saddle0-g} \\
\frac{\partial}{\partial \chi} \hat{W} & = & 
 m\tilde{\epsilon}- 2\upsilon \rho^{2} 
\frac{\gamma}{(\chi - \tau)^{3}}, 
    \label{saddle0-chi} \\
\frac{\partial}{\partial \tau} \hat{W} & = &  
\upsilon \rho^{2} \left[ m^{3}\Phi'(m\tau) + 
\frac{2\gamma}{(\chi - \tau)^{3}} \right]. 
    \label{saddle0-tau} 
\eea
as in Sect.~\ref{PT} let us denote the saddle point solutions 
for the dimensionless parameters $m\tau$, $m\chi$, $m\rho$ and $\gamma$ 
as $\ttau$, $\tchi$, $\trho$ and $\tgamma$ respectively. 
We are looking for these saddle point solutions as functions of  
$\tilde{\epsilon}$ and $\tilde{\nu}$. Here and below allowing 
an abuse of notation we will use the same letters for functions 
of $\teps$, $\tnu$ and $\epsilon$, $\nu$. This will not lead to a 
confusion. 

\begin{figure}[tp]
\begin{center}
\epsfig{file=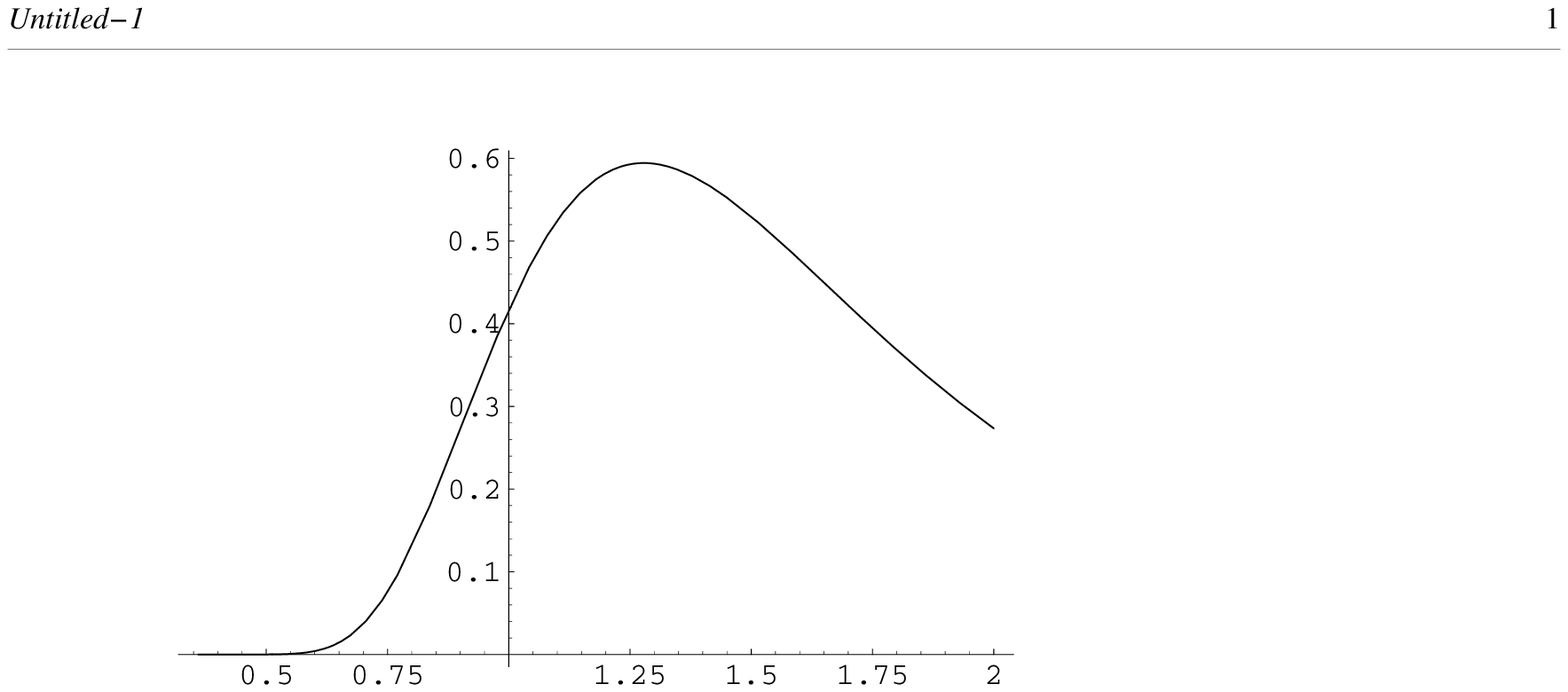,%
bbllx=100pt,bblly=530pt,%
bburx=490pt,bbury=745pt,%
clip=%
}
\end{center}
\caption{Plot of function $\epsilon (\ttau_{0})$ for $\nu = 0$.}
\label{fig:eps-tau0}
\end{figure}

\begin{figure}[tp]
\begin{center}
\epsfig{file=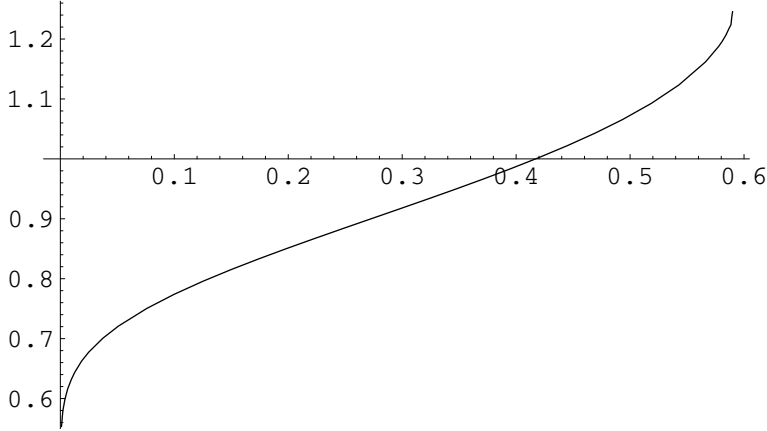,%
bbllx=100pt,bblly=550pt,%
bburx=490pt,bbury=745pt,%
clip=%
}
\end{center}
\caption{Plot of function $\ttau_{0} (\epsilon)$ for $\nu = 0$.}
\label{fig:tau-eps0}
\end{figure}

{}From Eqs. (\ref{saddle0-r2}) - (\ref{saddle0-tau}) in 
the leading order at $\nu \rightarrow 0$ we get 
\bea
\tilde{\rho}^{2} & = & - \frac{1}{\upsilon}
\frac{\tilde{\epsilon}}{\Phi'(\tilde{\tau})}, \label{sol-r2} \\
\tilde{\gamma} & = & - 4 \left( \frac{\tilde{\nu}}{\tilde{\epsilon}} 
\right)^{3} \Phi' (\tilde{\tau}), \label{sol-g} \\
\tilde{\chi} & = & \tilde{\tau} + 2\frac{\tilde{\nu}}{\tilde{\epsilon}}.   
\label{sol-x}
\eea
The function $\ttau(\teps,\tnu)$ is a solution of the equation 
\beq
\ln \left( - \frac{\tilde{\epsilon} C e}{\upsilon \Phi' (\tilde{\tau})}
  \right) + 4 \left[ \Phi (\tilde{\tau}) - 
\frac{\tilde{\nu}}{\tilde{\epsilon}} \Phi'(\tilde{\tau}) \right]  + 
{\cal O}(\nu^{2})= 0.   \label{tau-eqn}
\eeq
This equation is, of course, a consequence of the system of the saddle 
point equations (\ref{saddle0-r2}) - (\ref{saddle0-tau}). 
Its solution can be obtained as an expansion in $\tilde{\nu}$. We write 
\[
\ttau (\teps,\tnu) = \ttau_{0}(\teps) + \tnu \ttau_{1}(\teps) + 
{\cal O}(\tnu^{2}). 
\]
After substituting this expression into Eq. (\ref{tau-eqn}) one finds  
that the coefficient $\ttau_{1}$ is equal to 
\[
\ttau_{1} = \frac{4}{\teps} \frac{ \left(\Phi'(\ttau_{0})\right)}
{4 \left(\Phi'(\ttau_{0})\right) - \Phi''(\ttau_{0})}, 
\]
whereas the function $\ttau_{0}(\teps)$ is determined by the equation 
\beq
\ln \left( - \frac{\tilde{\epsilon} C e}
{\upsilon \Phi' (\tilde{\tau}_{0})}\right) + 
4 \Phi (\tilde{\tau}_{0})  = 0.   \label{tau0-eqn}
\eeq
Note that the solutions (\ref{sol-g}), (\ref{sol-x}) are in 
agreement with the behaviour (\ref{chi-tau-1}) assumed above. 
The behaviour of the solution $\ttau_{0}(\teps)$ can be analyzed 
by writing first equation (\ref{tau0-eqn}) as 
\[
\teps = - \frac{\upsilon}{Ce} \Phi' (\ttau_{0}) e^{-4\Phi (\ttau_{0})}, 
\]
studying the function $\teps (\ttau_{0})$ (or $\epsilon (\ttau_{0}) = 
\teps (\ttau_{0})/\kappa$) and then inverting it. The plots 
of functions  $\epsilon (\ttau_{0})$ and  $\ttau_{0} (\epsilon)$ are given in 
Fig.~\ref{fig:eps-tau0} and Fig.~\ref{fig:tau-eps0}.

Let us write the exponent $F(\epsilon,\nu)$ of  
the cross section (\ref{sigma-N}) as 
\[
F(\epsilon,\nu) = -32 \pi^{2} + F^{(1)}(\epsilon,\nu) + 
F^{(2)}(\epsilon,\nu) + \ldots, 
\]
where $F^{(1)}(\epsilon,\nu)$ and $F^{(2)}(\epsilon,\nu)$ are given 
by $\hat{W}$ and $W^{(2)}$, respectively, evaluated at the saddle point 
solution. Substituting solutions (\ref{sol-r2}) - (\ref{sol-x}) 
into Eq. (\ref{W1-3}) we obtain that 
\beq
F^{(1)}(\epsilon,\nu) = \kappa \epsilon \left[ \tilde{\tau}_{0}
(\epsilon) + \frac{1}{4 \Phi'(\tilde{\tau}_{0}(\epsilon))} \right]
+{\cal O}(\nu).              \label{F1-1}
\eeq

For $0 \leq \ttau_{0} \leq 1.28$ the function $\epsilon (\ttau_{0})$ 
grows monotonically from $\epsilon = 0$ to $\epsilon \approx 0.59$ 
(see Fig.~\ref{fig:eps-tau0}). Hence, for the interval $0 \leq \epsilon 
\leq 0.59$ there exists the inverse function $\ttau_{0}(\epsilon)$ 
with the property $\ttau_{0} (0) = 0$. This is also the interval 
of energies for which the saddle point solution at $\nu = 0$ 
can be found. The plot in Fig.~\ref{fig:tau-eps0} 
shows that for $0 \leq \epsilon \leq 1.59$ the values of 
$\ttau_{0}(\epsilon)$ in general are not small. Since $\ttau$ is 
the saddle point solution for $\tau m$, this observation suggests that 
the mass corrections cannot be neglected. For $0 < \epsilon < 0.1$ 
we enter the regime when $\ttau_{0} < 1$ and the expressions for 
the saddle point solution and Eq. (\ref{F1-1}) simplify considerably. 
In particular, $\Phi(\ttau_{0}) \sim 1/\ttau_{0}^{2}$ 
for $\ttau_{0} \ll 1$. In this regime 
corrections due to non-zero mass can be neglected. 

At very small energies, namely when 
\beq
  \frac{\ln \ln \frac{1}{\epsilon}}{\ln \frac{1}{\epsilon}} \ll 1, 
    \label{eps-small}
\eeq
we enter the regime when analytical expressions for the saddle point 
solutions can be obtained. The function $\ttau_{0}(\teps)$ 
is small in this limit. In the leading approximation 
Eq. (\ref{tau0-eqn}) takes the form 
\[
\ln \teps + \frac{4}{\ttau_{0}^{2}} + 5\ln \ttau_{0} + 
\ln \left( \frac{Ce}{2\upsilon} \right) + 2C_{E} -1 + \ldots = 0, 
\]
where the dots stand for terms vanishing as $\teps \rightarrow 0$. 
It is easy to obtain the asymptotic form of the solution of this equation in 
for the variable $1/\ttau_{0}(\teps)$. We get  
\[
\frac{1}{\ttau_{0}^{2}(\teps)} = \frac{1}{4} \ln \frac{1}{\teps} 
+ \frac{5}{8} \ln \ln \frac{1}{\teps} + \ln \upsilon -4C_{E} -1 + 
\; \; \mbox{decreasing terms}.   
\]
The first two leading terms of the function $\ttau_{0}(\teps)$  are 
given by 
\beq
  \tilde{\tau}_{0}(\teps)  = \frac{2}{\sqrt{\ln \frac{1}{\teps}}}
  - \frac{5}{2} \frac{\ln \ln \frac{1}{\teps}}
  {\left( \ln \frac{1}{\teps} \right)^{3/2}} + \ldots
\label{tau-1}
\eeq  
Then in the leading order in energy solutions 
(\ref{sol-r2}) - (\ref{sol-x}) become
\beq
\tilde{\rho}^{2} = \frac{1}{48 \pi^{2}}
\frac{\tilde{\epsilon}}{\left( 
\ln \frac{1}{\tilde{\epsilon}} \right)^{3/2}},  \quad
\tilde{\gamma} = \left( \frac{\tilde{\nu}}{\tilde{\epsilon}} \right)^{3}
\left( \ln \frac{1}{\tilde{\epsilon}} \right)^{3/2}, \quad
(\tilde{\chi} - \tilde{\tau}) = 
2 \frac{\tilde{\nu}}{\tilde{\epsilon}}.  \label{saddle-1}
\eeq   
In this regime the function $F^{(1)}(\epsilon, \nu)$ is equal to
\beq
F^{(1)}(\epsilon, \nu) = 2 \frac{\kappa \epsilon}
{\sqrt{ \ln \frac{1}{\epsilon}}} \left[ 1 + {\cal O}
\left( \frac{\ln \ln \frac{1}{\epsilon}}{\ln \frac{1}{\epsilon}}
\right) \right] + {\cal O}(\nu).    \label{F1}
\eeq 
This is result (\ref{F-LO}) quoted in the Introduction. 

The propagator correction $F^{(2)}(\epsilon,\nu)$ in the limit 
$\nu \rightarrow 0$ is obtained by evaluating contributions 
(\ref{W2}) - (\ref{W2-ii}) at the saddle point solution 
(\ref{sol-r2}) - (\ref{sol-x}). 
In Sect.~\ref{PT} we showed in a rather general context that singular 
terms $\ln (1/\nu)$ appear in the contributions involving initial 
particles and proved that they cancel each other in the final result. 
Let us discuss now the exact expression for the propagator correction for 
the $(-\lambda \phi^{4})$-model in the limit $\nu \rightarrow 0$. 
The result for the partial contributions is given by Eqs. 
(\ref{F2-ii-2}) - (\ref{F2-ff}) in the Appendix. We see that singular 
terms $\ln (1/\tnu)$ appear only in the functions 
$F^{(2)}_{(i-i)}(\epsilon,\nu)$ and of $F^{(2)}_{(i-f)}(\epsilon,\nu)$ 
which involve initial states and which are proportional 
to $\alpha_{1}$. Hence, according to Eq. (\ref{R1}), the singular terms  
are due to $s \ln s$- and $s$-terms in the instanton propagator. 
It is clear now that, if following the method of Ref. \cite{Mu92}, 
reviewed in Sect.~\ref{asympt}, one chooses the propagator constraint 
in such way that these terms are absent in the residue, 
then the contributions of the initial states are zero in the limit 
$\nu \rightarrow 0$. Correspondingly, in this case no singular terms 
appear. 

As it follows from these formulas and the 
saddle point solution (\ref{tau-1}), in the limit of small energies 
the corrections $F^{(2)}_{(i-i)}(\epsilon,\nu)$ and of 
$F^{(2)}_{(i-f)}(\epsilon,\nu)$ behave as 
\[
  \sim \frac{\epsilon^{3}}{\left( \ln \frac{1}{\epsilon} \right)^{3/2}}. 
\]
The contribution $F^{(2)}_{(f-f)}(\epsilon,\nu)$ of the final states 
contains a part $\sim \alpha_{1}$ and a part $\sim \alpha_{2}$. 
The former is due to $s \ln s$- and $s$-terms in the instanton 
propagator residue, the latter is due to  
$\ln s$-terms and $s$-independent terms of it (see Eq. (\ref{R1}).       
In the limit of small energies $F^{(2)}_{(f-f)}(\epsilon,\nu)$ 
behaves as $\epsilon^{2}$,  
thus giving the dominant contribution to the propagator correction. 

As an illustration let us demonstrate the cancellation of the 
singular terms explicitly in our model. Summing up 
corrections (\ref{F2-ii}) and (\ref{F2-if}) we get
\bea
& & F^{(2)}_{(i-i)}(\epsilon,\nu) + F^{(2)}_{(i-i)}(\epsilon,\nu) = 
32 \alpha_{1}\upsilon \trho^{6}  \nonumber \\
& & \times \left\{ - \frac{\tgamma}{2(\tchi - \ttau)^{3}} 
\left[ \frac{2\gamma}{(\tchi - \ttau)^{3}} + \Phi'(\ttau) \right] \cdot 
\left[ \frac{71}{30} + \ln \frac{\trho^{2}}{(\tchi - \ttau)^{2}} \right] 
  \right. \nonumber \\
& & \left. - \frac{\tgamma}{2(\tchi -\ttau)^{3}} \Phi'(\ttau) \ln 
\frac{\trho^{2}}{\ttau^{2}} + \frac{\tgamma}{2(\tchi - \ttau)^{3}} 
\Phi'(\ttau) \left(2\ln 2 - \frac{71}{30}\right) + 
\frac{\tgamma}{(\tchi - \ttau)^{3}}
{\cal A}_{1}(\ttau) \right\}. \nonumber 
\eea
The singular term is $\ln (\trho^{2}/(\tchi - \ttau)^{2}) \sim \ln (1/\tnu)$ 
in the second line. The coefficient in front of it vanishes 
exactly due to the saddle point equation (\ref{saddle0-tau}). 

Finally, let us present the result for the complete propagator correction 
to the exponent of the multiparticle cross section in the limit 
$\nu \rightarrow 0$. Summing up expressions (\ref{F2-ii}), 
(\ref{F2-if}) and (\ref{F2-ff}) for the partial corrections, we obtain 
\bea
F^{(2)}(\epsilon,0) & = & - \frac{\alpha_{2}}{\upsilon} \teps^{2} 
\frac{4\Phi^{2}(\ttau)}{\left( \Phi' (\ttau) \right)^{2}} 
\left[ \frac{43}{15} -\ln 2  - 2C_{E}- 4\Phi (\ttau) - 
\ln \frac{\ttau^{2}}{4} + \frac{{\cal A}_{A}(\ttau)}{\Phi^{2}(\ttau)} \right]     
            \nonumber   \\
 & - & \frac{8\alpha_{1}}{\upsilon^{2}}\frac{\teps^{3}}{\Phi'(\ttau)} 
\left\{ \frac{1}{2} - \ln 2 - 2 \frac{{\cal A}_{1}(\ttau)}{\Phi' (\ttau)} - 
\frac{{\cal A}_{3}(\ttau)}{\left( \Phi' (\ttau) \right)^{2}}\right. 
\label{F2-tot}  \\
 & + & \left. \frac{\Phi^{2}(\ttau)}{\left( \Phi' (\ttau) \right)^{2}}  
 \left[ - \ln 2 - \frac{2}{15} - 2C_{E} - 4\Phi(\ttau) - 
  2 \ln \frac{\ttau}{2} \right] \right\}.   \nonumber 
\eea   

The expression for the propagator correction simplifies 
in the regime $\ttau_{0} \ll 1$. In this case, using  
Eqs.(\ref{F2-ii-1}), (\ref{F2-if-1}) and (\ref{F2-ff-1}) or calculating 
the limit of small $\ttau_{0}$ directly from Eq. (\ref{F2-tot}), one gets 
\bea
F^{(2)}(\epsilon,0) & = & \frac{4\alpha_{2}}{\upsilon} \teps^{2} 
\left[ 1 + \frac{\ttau_{0}^{2}}{4} \left( 8 \ln \frac{\ttau_{0}}{2} + 8C_{E} 
- \frac{43}{15}\right) + {\cal O}(\ttau_{0}^{4}) \right] \label{F2-tot-1} \\
 & - & \frac{4\alpha_{1}}{\upsilon^{2}} ( \teps \ttau_{0})^{3} \cdot  
  {\cal O}\left(\ttau_{0}^{2},\ttau_{0}^{2}\ln \ttau_{0} \right). \nonumber 
\eea    
The first line is the leading term at small energies:
\beq
F^{(2)}(\epsilon,0) = \frac{4\alpha_{2}}{\upsilon} \teps^{2} = 
 \frac{1}{32\pi^{2}} \kappa^{2} \epsilon^{2}.  \label{F2-2}
\eeq
As it was discussed above, this is the contribution from the propagator 
between final-final state. The first correction to expression 
(\ref{F2-2}) is given by the $\ttau^{2} \ln \ttau$ 
term in Eq. (\ref{F2-tot-1}). In the regime of very small energies 
we have 
\[
   \ttau_{0}^{2} \ln \ttau_{0} \sim \frac{\ln \ln \frac{1}{\epsilon}}
   {\ln \frac{1}{\epsilon}}.
\]
By comparing the propagator correction (\ref{F2-2}) with the leading 
order correction (\ref{F1}) we see that the actual expansion parameter 
at small energies is indeed $\epsilon \sqrt{\ln (1/\epsilon)}$, as 
it was argued in Ref. \cite{Ti2}. 

\section{Propagator correction in general case}
\label{general}

\begin{figure}[tp]
\begin{center}
\epsfig{file=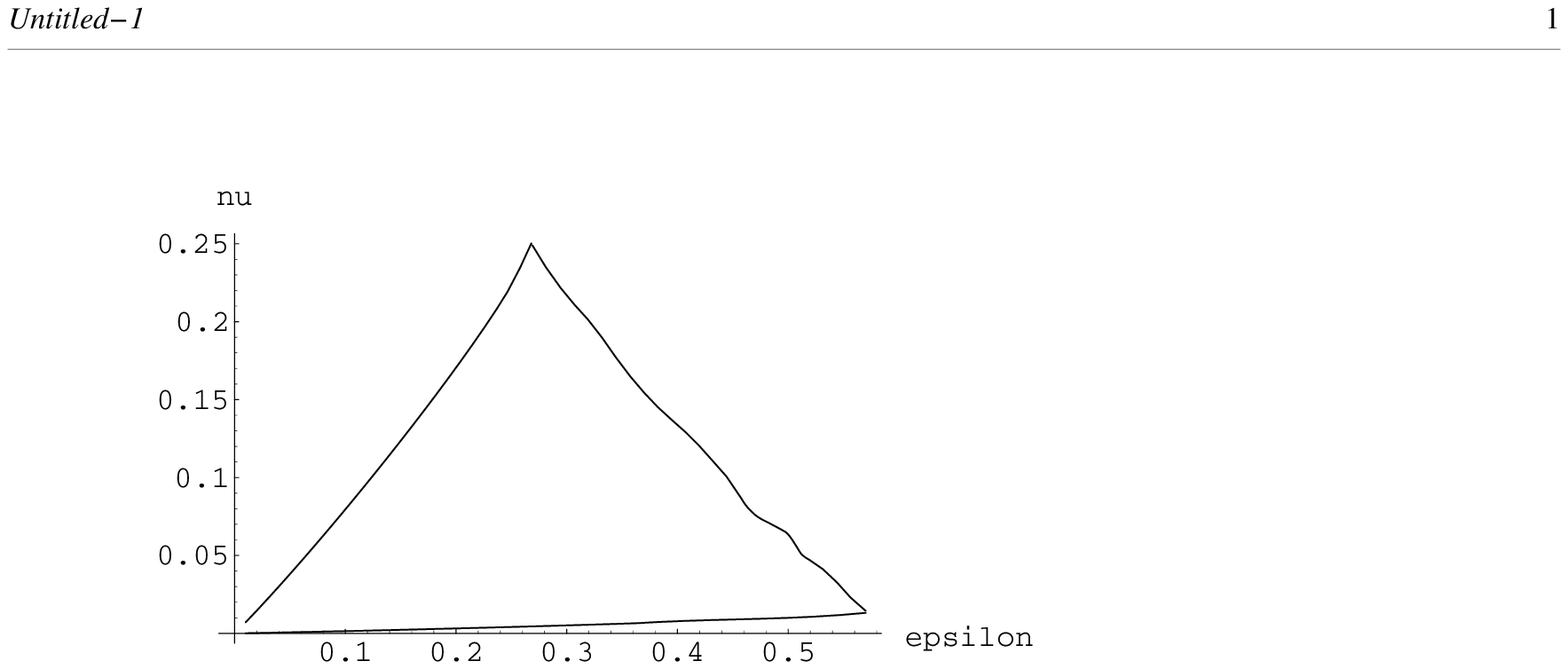,%
bbllx=100pt,bblly=550pt,%
bburx=490pt,bbury=745pt,%
clip=%
}
\end{center}
\caption{Region of values of $\epsilon$ and $\nu$ for which the saddle 
point solution exists.}
\label{fig:region}
\end{figure}

For arbitrary $\epsilon$ and $\nu$ the saddle point equations are 
obtained by calculating of the derivatives of the general 
expression $\hat{W}$, Eq. (\ref{W1-2}), with respect 
to $\rho^{2}$, $\gamma$, $\chi$ and $\tau$. In this case we found 
the saddle point solutions $\trho^{2}(\epsilon,\nu)$, $\ttau (\epsilon,\nu)$, 
$\tchi (\epsilon,\nu)$, $\tgamma (\epsilon,\nu)$ for $(m\rho)^{2}$, 
$m\tau$, $m\chi$, $\gamma$, respectively, and computed the functions 
$F^{(1)}(\epsilon,\nu)$, $F^{(2)}(\epsilon,\nu)$ 
numerically. It turned out that the saddle point solution 
exists only for the region in the $(\epsilon,\nu)$-plane which is 
presented in Fig.~\ref{fig:region}. It 
lies inside the rectangle $0 < \epsilon < \epsilon_{max}= 0.59$ and 
$0 < \nu < \nu_{max} = 0.25$. For points very close to the 
axis $\nu = 0$ our numerical computations fail. Extrapolating these 
numerical results to $\nu = 0$ we found good agreement with the 
analytical results for $\nu = 0$ obtained in Sect.~\ref{nu0}. 

We performed the numerical analysis for the whole region, plotted in 
Fig.~\ref{fig:region}.  
For the presentation of the results it is convenient to introduce the 
function ${\cal F}(\epsilon,\nu)$, related to $F(\epsilon,\nu)$ by 
\beq
  F(\epsilon,\nu) = - 32 \pi^{2} {\cal F}(\epsilon,\nu), \label{calF-def}
\eeq
and its leading and next-to-leading approximations: 
\[
{\cal F}_{1}(\epsilon,\nu) = 1 - \frac{F^{(1)}(\epsilon,\nu)}{32 \pi^{2}},
\; \; \mbox{and} \; \; 
{\cal F}_{2}(\epsilon,\nu) = 1 - \frac{F^{(1)}(\epsilon,\nu) + 
F^{(2)}(\epsilon,\nu)}{32 \pi^{2}}.
\]
All these functions are normalized by the conditions 
${\cal F}(0,\nu) = {\cal F}_{1}(0,\nu) = {\cal F}_{2}(0,\nu) = 1$. 

Before presenting our results we would like to mention that the 
complete function $F(\epsilon,\nu)$ in the range $0.4 < \epsilon < 3.5$ and 
$0.25 < \nu < 1$ was computed in Ref. \cite{KT}. 
The computation was performed by solving a certain classical 
boundary value problem on the lattice. With the size of the 
lattice used in the numerical calculation in Ref. \cite{KT}, 
the authors did not obtain data for smaller $\epsilon$ and $\nu$ 
except for the line of points corresponding to the periodic  
instanton solutions \cite{KRT-pi}. This line is directed from the zero 
energy instanton ($\epsilon = \nu =0$) to the sphaleron 
($\epsilon = \nu =1$).  

\begin{figure}[tp]
\epsfxsize=0.9\hsize
\epsfbox{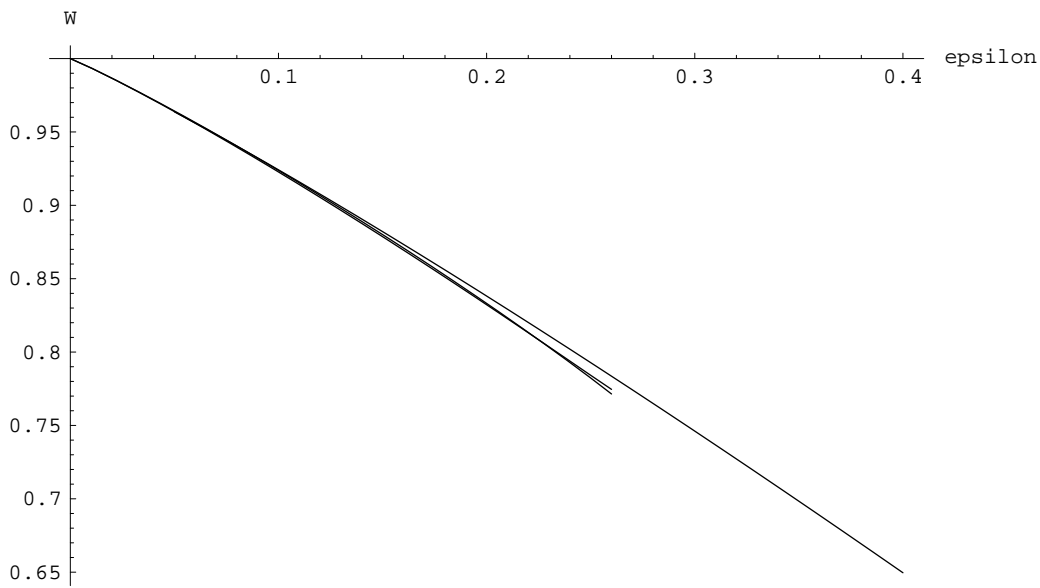}
\caption{Plots of ${\cal F}_{1}(\epsilon,\nu_{p}(\epsilon))$ 
(the lower curve) and ${\cal F}_{2}(\epsilon,\nu_{p}(\epsilon))$ 
as functions of $\epsilon$ for the periodic instanton. For comparison  
the complete function ${\cal F}(\epsilon,\nu_{p}(\epsilon))$ (the longer 
curve), calculated in Ref. \cite{KT}, is also plotted.}
\label{fig:periodic}
\end{figure}

The left upper bound of the region in Fig.~\ref{fig:region} is precisely 
the line $\nu=\nu_{p}(\epsilon)$ of values of 
$\epsilon$ and $\nu$ of periodic instantons. For them 
$\tilde{\tau}(\epsilon,\nu) = \tilde{\chi}(\epsilon,\nu)/2$. 
The leading order and the next-to-leading order 
approximations for ${\cal F}(\epsilon,\nu)$ at the line 
$(\epsilon,\nu_{p}(\epsilon))$ are shown
in Fig.~\ref{fig:periodic}. One can see that in the whole range 
of calculation these curves are close to each other. 
They can be compared with the complete function 
${\cal F}(\epsilon,\nu_{p}(\epsilon))$ for the line of periodic instantons,  
computed numerically in Ref. \cite{KT} and also plotted in 
Fig.~\ref{fig:periodic}. The comparison shows that our perturbative results 
do not differ significantly from the exact ones  
for $\epsilon < 0.25$ and $\nu < 0.2$. These values can be regarded as a 
rough estimate of the range of validity of the perturbative calculations up 
to the next-to-leading order. 

\begin{figure}[tp]
\epsfxsize=0.9\hsize
\epsfbox{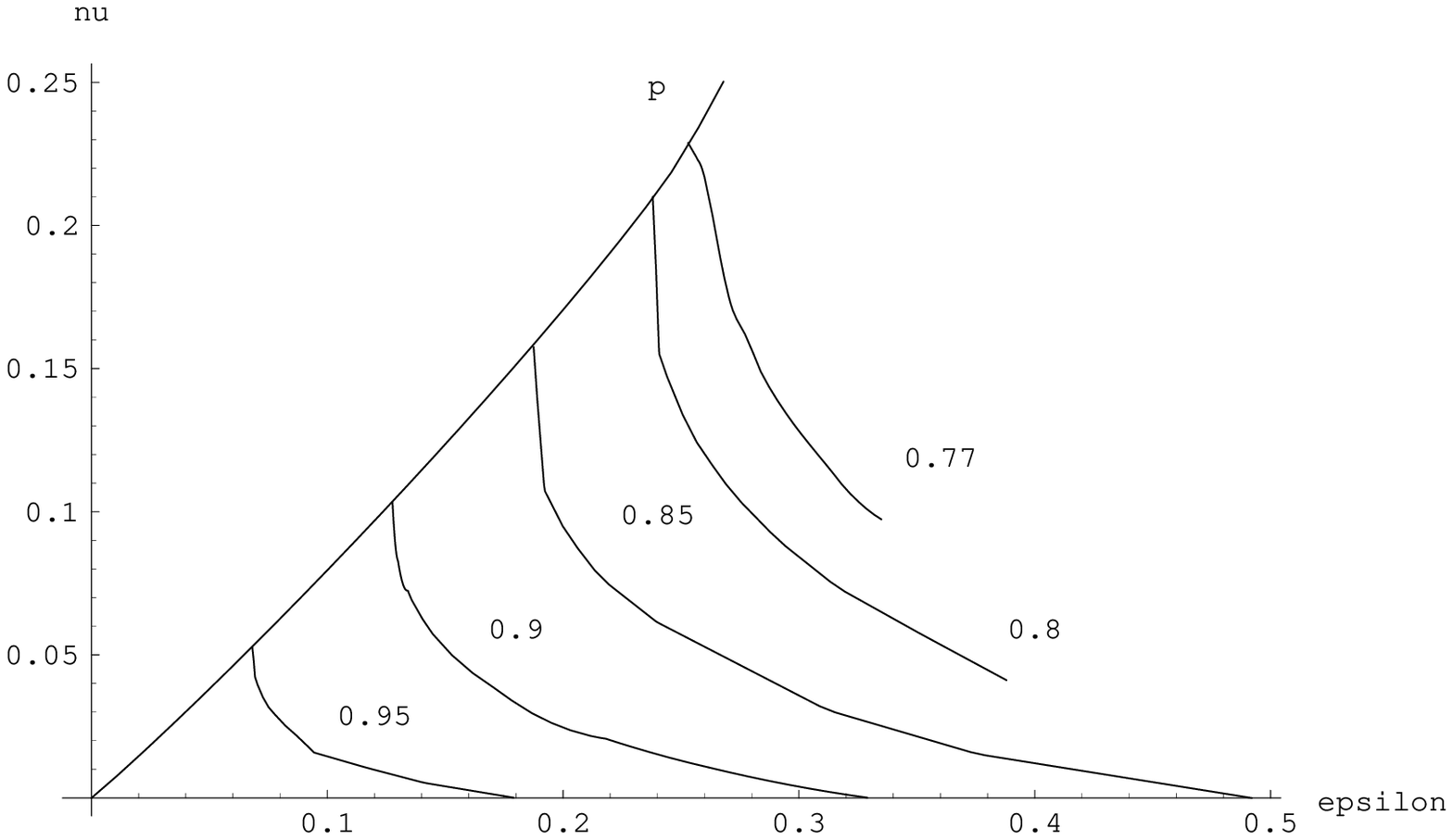}
\caption{Lines of constant ${\cal F}_{2}(\epsilon,\nu)$ in the 
$(\epsilon,\nu)$-plane. Numbers near the lines indicate the 
value of ${\cal F}_{2}$, ``p'' labels the line of periodic instanton 
solutions.}
\label{fig:Fconst}
\end{figure}

Lines of constant ${\cal F}_{2}(\epsilon,\nu)$ are plotted in 
Fig.~\ref{fig:Fconst}. Within the range of $(\epsilon,\nu)$ in 
Fig.~\ref{fig:region} the minimal value of ${\cal F}(\epsilon,\nu)$ 
is a bit below ${\cal F}=0.77$. This gives still a considerable 
suppression of the multiparticle cross section of instanton 
induced processes. 

As it was explained above, our main interest is 
to study the cross section for shadow processes with a 
few initial particles. According to conjecture (\ref{conj}) 
the points, where the lines of constant ${\cal F}_{2}$ 
cross the $\nu = 0$ axis, are of particular 
interest. Extrapolating our numerical results to $\nu =0$, one obtains, 
for example, ${\cal F}_{2}(\epsilon,0)=0.95$ at 
$\epsilon = 0.180$, ${\cal F}_{2}(\epsilon,0)=0.85$ at 
$\epsilon = 0.492$. These values, of course, can be obtained 
by using the exact formulas for $\nu = 0$ from Sect.~\ref{nu0}. 
We would like to mention that in the region of $(\epsilon,\nu)$ which 
we studied the next-to-leading approximation ${\cal F}_{2}(\epsilon,\nu)$ is 
quite small comparing to the leading order ${\cal F}_{1}(\epsilon,\nu)$. 
The difference between these functions does not exceed $10^{-2}$. 
The curves in Fig.~\ref{fig:Fconst} end at the line formed by saddle points 
of the periodic instanton solutions.

In Sect.~\ref{residue} we mentioned that various expressions for the 
functions $s_{\#}(\bk,\bq)$ can be used in the definition of 
the residue of the instanton propagator, Eq. (\ref{RR1}). 
Examples of such functions are given by Eqs. (\ref{saa}), (\ref{sab}) 
and (\ref{tsaa}), (\ref{tsab}). 
We performed the calculation of the leading and next-to-leading order 
approximations for these particular choices of functions and for 
a wide range of values of $\epsilon$ and $\nu$. We found that the 
difference is very tiny. Thus, 
\[
{\cal F}_{2}(0.2,\nu_{p}(0.2))|_{\tilde{s}_{aa},\tilde{s}_{ab}} - 
{\cal F}_{2}(0.2,\nu_{p}(0.2))|_{s_{aa},s_{ab}} \approx 5 \cdot 10^{-4}.
\]

\section{Accuracy of the approximation}
\label{approx}

As it was explained in Sect.~\ref{model}, in principle the constrained 
instanton configuration and the action of this configuration 
depend on the constraints imposed. In Eq. (\ref{S-ci}) we calculated 
only the leading, constraint independent part of the action. This is 
main restriction to the accuracy which can be achieved in our calculations. 
Corrections due to the form of the constraint after evaluation at 
the saddle point solution are of the order $\trho^{4}$, 
$\trho^{2} \ln \trho$ or $\trho^{2} \ln^{2} \trho$ (see Eq. (\ref{S-ci})).  
It turns out that within this accuracy main terms of the leading 
and propagator contributions to $F(\epsilon,\nu)$ can be calculated. For 
this one has to follow a certain scheme of approximations, which 
we discuss in the present section.  

Let us first obtain some simple estimates of the leading and 
next-to-leading corrections at $\nu \rightarrow 0$. For the case 
when $m\tau \ll 1$ the dominating term of the leading correction is 
given by $\rho^{2} J_{0}(\gamma,\tau,\chi) \sim \rho^{2}/\tau^{2}$. 
{}From the analysis of the next-to-leading one can show that the 
main term is $F_{(f-f)}^{(2)}$, which is due to the instanton 
propagators between the final states. For small $m\tau$ the dominant 
contribution to the integrals in Eq. (\ref{W2-ff}) are given by 
the soft momenta $|\bk| \sim |\bq| \sim 1/\tau$. For such momenta the 
$\ln (\rho^{2}s)$-term in the expression (\ref{R1}) for the residue of the 
instanton propagator plays the main role. Putting all the factors 
together one can see that $F^{(2)} \sim F^{(2)}_{(f-f)} \sim 
(\rho/\tau)^{4} \ln (\rho/\tau)^{2}$. To summarize, in this regime 
\beq
F^{(1)}(\epsilon,\nu) \sim \frac{\rho^{2}}{\tau^{2}}, \; \; \; 
F^{(2)}(\epsilon,\nu) \sim\frac{\rho^{4}}{\tau^{4}} 
\ln \frac{\rho^{2}}{\tau^{2}}. \label{F1F2-est}
\eeq
These estimates are in agreement with results (\ref{F1-1}), (\ref{F2-tot-1}) 
and (\ref{F2-ff-2}). To see this one should use formula (\ref{sol-r2}) for 
$\trho^{2}$ and take into account that for $\ttau \ll 1$ 
\[
\teps \ttau \sim \trho^{2} \ttau \Phi'(\ttau) \sim \frac{\trho^{2}}{\ttau^{2}} 
\]
in Eq. (\ref{F1-1}), and 
\[
\trho^{4} \Phi^{2}(\ttau) \ln \frac{\trho^{2}}{\ttau^{2}} \sim 
\frac{\trho^{4}}{\ttau^{4}} \ln \frac{\trho^{2}}{\ttau^{2}}
\]
in Eq. (\ref{F2-ff-2}). For very small energies from solutions 
(\ref{tau-1}), (\ref{saddle-1}) one gets 
\beq
\frac{\trho^{2}}{\ttau^{2}} \sim 
\frac{\epsilon}{\sqrt{\ln \frac{1}{\epsilon}}}, \; \; \;
\frac{\trho^{4}}{\ttau^{4}} \ln \frac{\trho^{2}}{\ttau^{2}} \sim 
\epsilon^{2}   \label{estimates}
\eeq
in accordance with results (\ref{F1}) and (\ref{F2-2}). 

\begin{figure}[tp]
\begin{center}
\epsfig{file=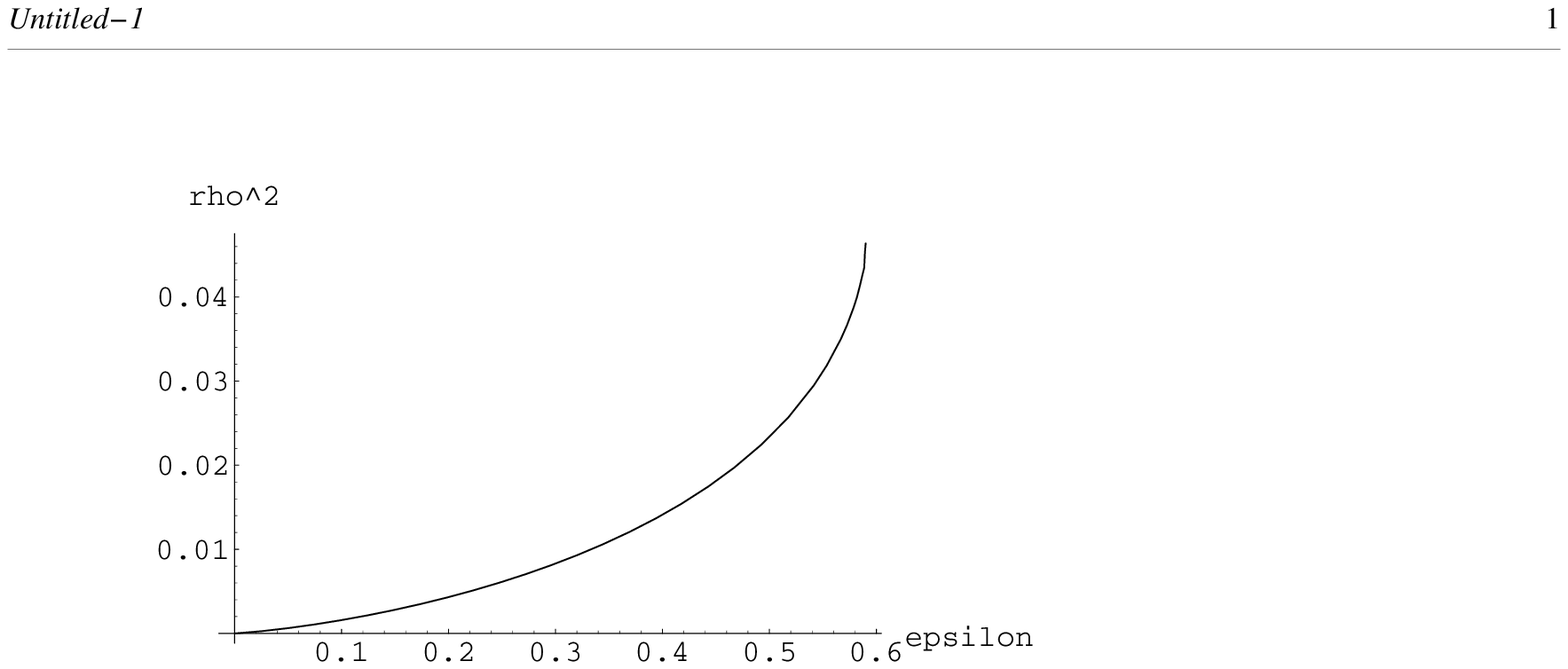,%
bbllx=100pt,bblly=550pt,%
bburx=490pt,bbury=745pt,%
clip=%
}
\end{center}
\caption{Plot of the function $\trho^{2}(\epsilon,0)$.}
\label{fig:r2}
\end{figure}

\begin{figure}[tp]
\begin{center}
\epsfig{file=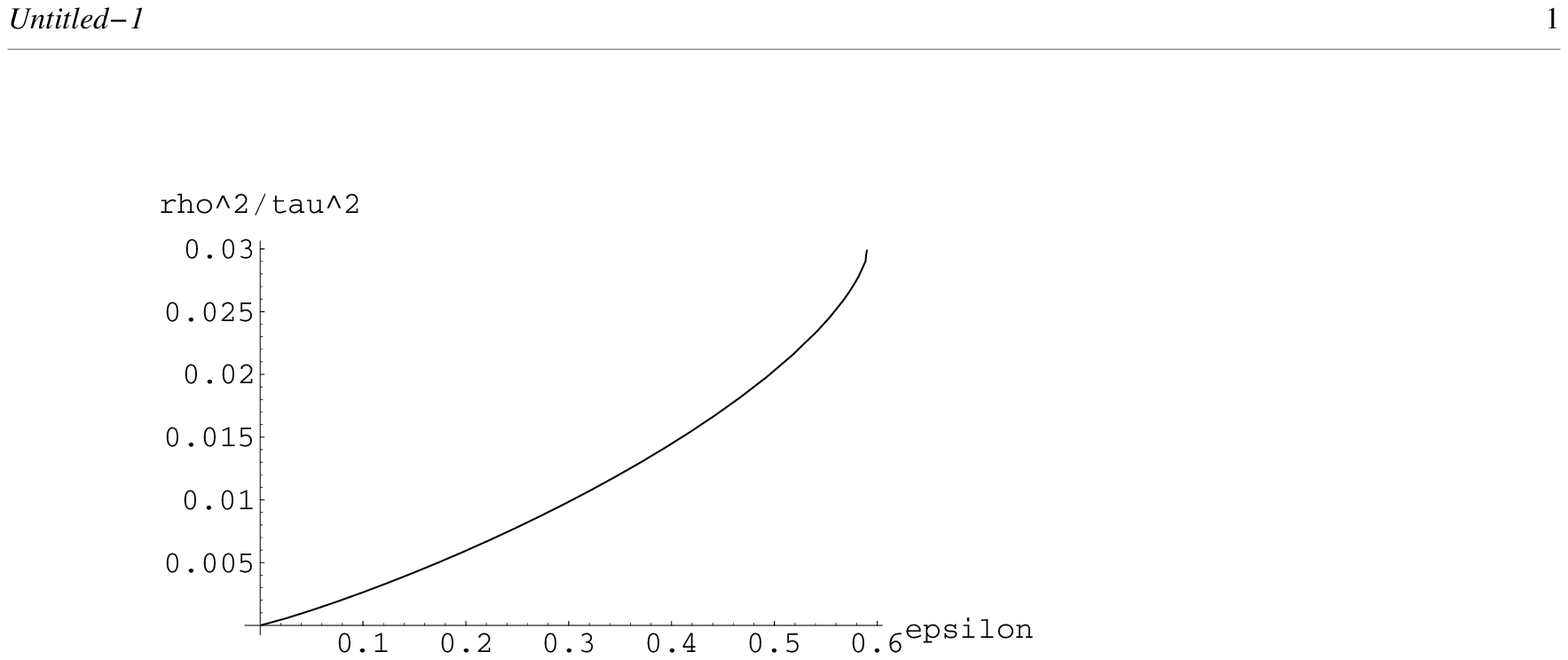,%
bbllx=100pt,bblly=550pt,%
bburx=490pt,bbury=745pt,%
clip=%
}
\end{center}
\caption{Plot of the function 
$\trho^{2}(\epsilon,0)/\ttau^{2}(\epsilon,0)$.}
\label{fig:r2/t2}
\end{figure}

To have numerical estimates of the range of values of these characteristic 
terms we studied the saddle point solution $\trho^{2}(\epsilon,\nu)$ and 
the ratio $\trho^{2}(\epsilon,\nu)/\ttau^{2}(\epsilon,\nu)$ numerically 
for a wide range of $\epsilon$ and $\nu$. In Fig.~\ref{fig:r2} and 
Fig.~\ref{fig:r2/t2} we present the plots of these functions for the 
most interesting case $\nu = 0$.   

For the procedure of calculation to be consistent one has to check, 
first of all, that $\trho^{4}$, $\trho^{4} \ln \trho$ and 
$\trho^{4} \ln^{2} \trho$ are smaller than the propagator correction 
$F^{(2)}(\epsilon,\nu)$. We checked numerically 
that this is indeed the case for 
a wide range of $\epsilon$ and $\nu$. In general, since, as it was 
mentioned in Sect.~\ref{general}, $|{\cal F}_{1} - {\cal F}_{2}| \leq 
10^{-3}$, the propagator correction $F^{(2)} \leq 0.3$, 
whereas $\trho^{4} \leq 10^{-4}$. Thus, at the point $\epsilon = 0.2$, 
$\nu_{p}(0.2)=0.17$ on the line of periodic instantons 
$F^{(2)}(0.2,0.17) = 0.28$, whereas 
$\trho^{2}(0.2,0.17) \approx 0.02$. Another example: for $\nu = 0$ 
$F^{(2)}(0.33,0) \approx 0.3$, whereas 
$\trho^{2}(0.33,0) \approx 0.01$. From Fig.~\ref{fig:r2}) and 
Fig.~\ref{fig:r2/t2} one can see that the corrections $\trho^{4}$, 
$\trho^{4} \ln \trho$ and $\trho^{4} \ln^{2} \trho$ are smaller than 
$F^{(2)}(\epsilon,\nu)$ for the regime when $\ttau < 1$ 
(see Eq. (\ref{F1F2-est}). This conclusion is also confirmed 
for $\nu =0$ in the limit of very small energies. Using formulas 
(\ref{tau-1}), (\ref{saddle-1}) one can see that 
\[
\trho^{4} \sim \frac{\epsilon^{2}}{\left( \ln \frac{1}{\epsilon} 
\right)^{3}}, \; \; \; 
\trho^{4} \ln \trho \sim  
\frac{\epsilon^{2}}{\left( \ln \frac{1}{\epsilon} \right)^{2}}, \; \; \; 
\trho^{4} \ln^{2} \trho \sim  
\frac{\epsilon^{2}}{\ln \frac{1}{\epsilon}}.   
\]
Recall that the propagator correction 
$F^{(2)}(\epsilon,0) \sim \epsilon^{2}$ in this regime 
(see Eqs. (\ref{F2-2}) and (\ref{estimates})). To give an illustration we 
presented the plot of the function $\trho (\epsilon, 0)$ in 
Fig.~\ref{fig:r2}. 

It turns out that with the accuracy set above it is enough to 
use the residue of the instanton solution $I(\rho)$ and the 
double on-mass-shell residue of the instanton propagator 
$P(\bk,\bq)$ of the massless theory. Let us justify this 
point. 

Let $\phi_{c.i.}(x;\rho)$ be the constrained instanton solution 
of the size $\rho$ in the massive theory. From general arguments 
one can see that its Fourier transform can be written as 
\[
\tilde{\phi}_{c.i.}(k;0,\rho) = \rho \frac{g(k^{2}\rho^{2},m^{2}\rho^{2})}
{k^{2} + m^{2}}. 
\]
Expanding the function $g(k^{2}\rho^{2},m^{2}\rho^{2})$ in powers of 
$m^{2}$ we write 
\[
g(k^{2}\rho^{2},m^{2}\rho^{2}) = g_{0}(k^{2}\rho^{2}) + 
(\rho m)^{2} g_{2}(k^{2}\rho^{2}) + \ldots
\]
Logarithmic terms of the form $(\rho m)^{2} \ln (\rho m)$ may 
also appear, however we will not write them explicitly assuming 
that they are roughly of the order $(\rho m)^{2}$. 
The residue $I_{c.i.}(\rho) = 
(k^{2}+m^{2}) \tilde{\phi}_{c.i.}(k;\rho)|_{k_{0}=iw_{\bk}}$ is 
equal to 
\[
I_{c.i.}(\rho) = \rho g_{0}(0) + \rho (\rho m)^{2} 
\left( g_{2}(0) - g_{0}'(0) \right) + \ldots
\]
Since the constrained instanton solution reproduces the instanton 
solution $\phi_{inst}(x;0,\rho)$, Eq. (\ref{inst}), of the massless 
theory in the limit $m \rightarrow 0$, the 
Fourier transform of $\phi_{inst}$ and its residue can be written as 
\[
\tilde{\phi}_{inst}(k;0,\rho) = \rho \frac{g_{0}(k^{2}\rho^{2})}{k^{2}}, 
\; \; \; I(\rho) = \rho g_{0}(0). 
\]
These formulas are in accordance with exact expressions (\ref{inst-FT}) 
and (\ref{R-inst}). 

Typical terms of the leading order correction are of the form 
\beq
\int \frac{d\bk}{w_{\bk}}  
I_{c.i.}(\rho) K(w_{\bk}\tau,w_{\bk}\chi) I_{c.i.}(\rho),
\label{LO-typ}
\eeq
where
\beq
K(w_{\bk}\tau,w_{\bk}\chi) = 
\frac{e^{-w_{\bk}\tau}}{1-\gamma e^{-w_{\bk}\chi}} \label{K-def}
\eeq
(see Eq. (\ref{W1-2})). By a simple analysis one can show that 
\bea
& & \int \frac{d\bk}{w_{\bk}}  
I_{c.i.}(\rho) K(w_{\bk}\tau,w_{\bk}\chi) I_{c.i.}(\rho) \nonumber \\
 & & = \int \frac{d\bk}{w_{\bk}}  
I_{inst}(\rho) K(w_{\bk}\tau,w_{\bk}\chi) I_{inst}(\rho) +   
 {\cal O} \left( m\rho^{2} \frac{\rho^{2}}{\tau^{2}}\right). \nonumber 
\eea
We checked that the ratio $\trho^{4}/\ttau^{2}$ is small comparing to 
$F^{(2)}(\epsilon,\nu)$ for a wide range of $\epsilon$ and $\nu$. 
This can also be seen from the plots in Fig.~\ref{fig:r2}) and 
Fig.~\ref{fig:r2/t2}. In the regime of very small energies from 
Eqs. (\ref{tau-1}), (\ref{saddle-1}) one gets 
\[
 \trho^{2} \frac{\trho^{2}}{\ttau^{2}} \sim 
 \frac{\epsilon^{2}}{\left( \ln \frac{1}{\epsilon} \right)^{2}}, 
\]
whereas $F^{(2)}(\epsilon,0) \sim \epsilon^{2}$. 

However, if the expression $w_{\bk}=|\bk|$ of the massless theory 
is used in (\ref{LO-typ}), then by simple estimates one can see that 
\beq
  \int \frac{d\bk}{w_{\bk}} I(\rho) K(w_{\bk}\tau,w_{\bk}\chi) 
  I(\rho) \sim 
  \int \frac{d\bk}{|\bk|} I(\rho) K(|\bk|\tau,|\bk|\chi) I(\rho) + 
\frac{\rho^{2}}{\tau^{2}} \cdot {\cal O}(m^{2}\tau^{2}).  \label{LO-typ1}
\eeq
Since the saddle point solution $\ttau (\epsilon,\nu) \geq  1$ for 
a range of values of $\epsilon$ and $\nu$ (see, for example, 
Fig.~\ref{fig:tau-eps0}), relation (\ref{LO-typ1}) warns that the 
corrections due to non-zero mass cannot be discarded. In particular, for the 
regime of very small energies and $\nu =0$ the term 
$(\trho^{2}/\ttau^{2}) \ttau^{2} \sim \epsilon / \ln (1/\epsilon)$ 
and obviously exceeds the propagator correction $F^{(2)} \sim \epsilon^{2}$. 
The conclusion is the following: in order to be within the accuracy set 
above one has to use the expression $w_{\bk}=\sqrt{\bk^{2}+m^{2}}$ 
for the energy, though it is enough to use the residue $I(\rho)$ 
of the instanton solution of the massless theory. 

For the residue of the instanton propagator and the propagator correction 
a similar analysis can be carried out. A general expression for 
the Fourier transform of the instanton propagator in the massive theory 
is of the form 
\[
G_{\psi}(k,q) = \rho^{2} \frac{h(k^{2}\rho^{2}, q^{2}\rho^{2}, s\rho^{2}, 
m^{2}\rho^{2})}{(k^{2}+m^{2})(q^{2}+m^{2})},
\]
where $s=(k+q)^{2}$. Expanding the function in the numerator 
in powers of $m^{2}\rho^{2}$, 
\[
h(k^{2}\rho^{2}, q^{2}\rho^{2}, s\rho^{2}, m^{2}\rho^{2}) = 
h_{0}(k^{2}\rho^{2}, q^{2}\rho^{2}, s\rho^{2}) + (\rho m)^{2} 
h_{2}(k^{2}\rho^{2}, q^{2}\rho^{2}, s\rho^{2}) + \ldots
\]
(logarithmic terms, like $\rho^{2}m^{2} \ln (\rho^{2}m^{2})$, may 
also appear), we obtain that the double on-mass-shell residue of the 
propagator can be written as 
\bea
P^{(m)}(s_{\#}\rho^{2}) & = & \rho^{2} h_{0}(0,0,s_{\#}\rho^{2}) + 
\rho^{2} (\rho m)^{2} \left[ h_{2}(0,0,s_{\#}\rho^{2}) \right. \nonumber \\
 & - & \left. h_{01}'(0,0,s_{\#}\rho^{2}) - h_{02}'(0,0,s_{\#}\rho^{2}) 
\right],  \nonumber 
\eea
where $\# = aa,ab,bb$, $s_{\#}$ is given by Eqs. (\ref{saa}), (\ref{sab})
and the upper index $(m)$ refers to the non-zero mass case. 
It is easy to see that the residue of the instanton propagator in the 
massless theory is then given by 
\[
P(s\rho^{2}) = \rho^{2} h_{0}(0,0,s\rho^{2}).
\]
A typical term in the propagator correction is 
\[
\rho^{2}\int \frac{d\bk}{w_{\bk}} \frac{d\bq}{w_{\bq}} I_{c.i.} 
K(w_{\bk}\tau_{1},w_{\bk}\chi) P^{(m)}(\rho^{2}s_{\#}(\bk,\bq)) 
K(w_{\bq}\tau_{2},w_{\bq}\chi) I_{c.i.}, 
\]
where $K(w_{\bk}\tau_{1},w_{\bq}\chi)$ is defined by Eq. (\ref{K-def}). 
By a simple analysis one gets 
\bea
& & \rho^{2}\int \frac{d\bk}{w_{\bk}} \frac{d\bq}{w_{\bq}} I_{c.i.}  
K(w_{\bk}\tau_{1},w_{\bk}\chi) P^{(m)}(\rho^{2}s_{\#}(\bk,\bq)) 
K(w_{\bq}\tau_{2},w_{\bq}\chi) I_{c.i.}   \nonumber \\
& & =\rho^{2} \int \frac{d\bk}{w_{\bk}} \frac{d\bq}{w_{\bq}} I  
K(w_{\bk}\tau_{1},w_{\bk}\chi) P(\rho^{2}s_{\#}(\bk,\bq)) 
K(w_{\bq}\tau_{2},w_{\bq}\chi) I \nonumber \\
& & + {\cal O} \left( \frac{\rho^{4}}{\tau^{4}} (\rho m)^{2}, 
(\rho m)^{6} \right).  \nonumber  
\eea
Again, from our numerical results we see that the corrections are 
small comparing to the propagator correction $F^{(2)}(\epsilon,\nu)$, 
calculated with $I$ and $P(\rho^{2}s)$, and can be neglected without 
loss of accuracy. Similar to estimate (\ref{LO-typ1}), if instead of 
$w_{\bk}=\sqrt{\bk^{2}+m^{2}}$ and $s_{\#}(\bk,\bq)$ the expressions 
$|\bk|$ and $s_{\#}^{(0)}(\bk,\bq)$, Eq. (\ref{s-0}), respectively, 
are used, then one can easily show that 
\bea
& & \rho^{2} \int \frac{d\bk}{w_{\bk}} \frac{d\bq}{w_{\bq}} I  
K(w_{\bk}\tau_{1},w_{\bk}\chi) P(\rho^{2}s_{\#}(\bk,\bq)) 
K(w_{\bq}\tau_{2},w_{\bq}\chi) I \nonumber \\ 
& & = 
\rho^{2} \int \frac{d\bk}{|\bk|} \frac{d\bq}{|\bq|} I  
K(|\bk|\tau_{1},|\bk|\chi) P(\rho^{2}s_{\#}^{(0)}(\bk,\bq)) 
K(|\bq|\tau_{2},|\bq|\chi) I\nonumber \\ 
& & + \left( \frac{\rho}{\tau} \right)^{4} {\cal O} \left( (m\tau)^{2}, 
(m\rho)^{2} \ln \frac{\rho^{2}}{\tau^{2}}, (m\rho)^{2} \right), 
\label{NLO-typ}
\eea
where we assumed that $\tau \sim \tau_{1} \sim \tau_{2}$. 
Since for a certain range 
of $\epsilon$ and $\nu$ the saddle point value $\ttau \sim 1$, 
Eq. (\ref{NLO-typ}) gives an indication that the mass corrections 
can be comparable to the propagator correction 
$F^{(2)} \sim (\trho/\ttau)^{4} \ln (\trho^{2}/\ttau^{2})$ 
and should be taken into account. 

Now we can formulate a procedure of calculation 
of the leading and next-to-leading order corrections to the function 
$F (\epsilon,\nu)$ as follows: in expressions for $\hat{W}$ and $W^{(2)}$

\begin{description}
\item[1)]  the residue 
of the instanton solution $I(\rho)$, given by Eq. (\ref{R-inst}), and 
the residue of the instanton propagator $P(\rho^{2}s)$, given by 
Eq. (\ref{R}), of the {\it massless} theory are substituted; 

\item[2)] the energy $w_{\bk}$ and the $s$-variables $s_{aa}$, 
$s_{ab}$ and $s_{bb}$, given by Eqs. (\ref{saa}), (\ref{sab}) (or by 
by Eqs. (\ref{tsaa}), (\ref{tsab})) of the {\it massive} theory 
are used. 
\end{description}

We have shown that this procedure is consistent and 
gives results which are within the accuracy, set by Eq. (\ref{S-ci}), 
provided $\trho^{2}(\epsilon,\nu)$ and $\trho^{2}(\epsilon,\nu)/ 
\ttau^{2}(\epsilon,\nu)$ are small in the range of our calculations. 
{}From the results of our numerical computations in Sect.~\ref{general}
we checked that this condition holds. 

\section{Discussion and conclusions}
\label{conclusion}

In the present paper we have analyzed the multiparticle cross section
of the shadow processes induced by instanton transitions in the simple
scalar model with the action given by Eq. (\ref{action}). 
We calculated the exact analytical
expression for the on-shell residue of the propagator of quantum
fluctuations in the instanton background. Using this result we calculated 
the propagator correction (i.e. the next-to-leading order correction) 
to the suppression factor $F/\lambda$ of the multiparticle cross section. 

The leading and next-to-leading (propagator) corrections to 
the function $F(\epsilon,\nu)$ were calculated semiclassically, 
by evaluation of the leading and next-to-leading terms in Eq. 
(\ref{sigma-N2}), respectively, at the saddle point. 

Within the accuracy of the approximation, considered in this 
article, the saddle point equations are derived from the leading 
terms, Eqs. (\ref{Gamma}), (\ref{Gamma-1}), in the 
"effective action". It turned out, that for such equations the saddle 
point solutions for the dimensionless parameters $(m\rho)^{2}$, 
$(m\tau)^{2}$, $(m\chi)^{2}$ and $\gamma$ in the integral 
(\ref{sigma-N2}) exist only for a certain region of 
$\epsilon = E/E_{sph}$ and $\nu = N/N_{sph}$, shown in 
Fig.~\ref{fig:region}. 

For general $\epsilon$ and $\nu$ from this region we computed 
the leading and propagator corrections to the function $F(\epsilon,\nu)$ 
numerically. For $\nu = 0$ we derived the analytical expressions 
for $F^{(1)}(\epsilon,0)$ and $F^{(2)}(\epsilon,0)$, Eqs. (\ref{F1-1}), 
(\ref{F2-ii-2}) - (\ref{F2-ff}), in terms of the 
saddle point solution $\ttau_{0}(\epsilon)$ of Eq. (\ref{tau0-eqn}). 
For the regime of very small energies, namely when relation 
(\ref{eps-small}) holds, and $\nu = 0$ we obtained explicit expressions 
for the saddle point solutions and the leading and propagator corrections 
to $F(\epsilon,0)$. Recall that according to conjecture (\ref{conj}) 
this is the function $F(\epsilon,0)$ which characterizes the cross section 
of the instanton induced processes with a few initial particles. 

The corrections $F^{(1)}(\epsilon,\nu)$  and $F^{(2)}(\epsilon,\nu)$ 
approach zero value when $\epsilon \rightarrow 0$. However, apparently 
there is no clear expansion parameter. At $\nu =0$ one can 
see that at energies for which $\ttau (\epsilon) \ll 1$ 
\bea
F^{(1)}(\epsilon,0) & = & 2 \upsilon \left( \frac{\trho^{2}}{\ttau^{2}}
\right)  + \ldots,            \label{F1F2-0} \\
F^{(2)}(\epsilon,0) & = & F^{(2)}_{(f-f)}(\epsilon,0) + \ldots 
= - 4 \alpha_{2} \upsilon \left( \frac{\trho^{4}}{\ttau^{4}} \right) 
\ln \frac{\trho^{2}}{\ttau^{2}} + \ldots. \nonumber 
\eea
Hence, the combination $(\trho^{2}/\ttau^{2}) \ln (\trho^{2}/\ttau^{2})$ 
can play the role of the small expansion parameter. For very small 
energies from Eqs. (\ref{tau-1}), (\ref{saddle-1}) we obtain that 
\bea
\frac{\trho^{2}(\epsilon,0)}{\ttau_{0}^{2}(\epsilon)} & = & 
\frac{1}{\upsilon} \frac{\teps}{\sqrt{\ln \frac{1}{\teps}}}
\left[ 1 + {\cal O} \left(\frac{\ln \ln (1/\teps)}{\ln (1/\teps)} 
\right) \right], \nonumber \\
\frac{\trho^{4}(\epsilon,0)}{\ttau_{0}^{4}(\epsilon)}
  \ln \frac{\trho^{2}(\epsilon,0)}{\ttau_{0}^{2}(\epsilon)} & = & 
  - \frac{\teps^{2}}{\upsilon^{2}}
  \left[ 1 + {\cal O} \left(\frac{\ln \ln (1/\teps)}{\ln (1/\teps)} 
  \right) \right].   \nonumber
\eea
We see that in this regime the expansion parameter is 
$\epsilon \sqrt{\ln (1/\epsilon)}$. With these relations, of course, 
results (\ref{F1}) and (\ref{F2-2}) are easily recovered from 
(\ref{F1F2-0}) (recall that $\upsilon = 192 \pi^{2}$ and $\alpha_{2}=3/2$). 

The range of validity of the next-to-leading order of approximation 
of the function $F(\epsilon,\nu)$ was estimated by comparing 
our results with numerical computations of the function 
$F(\epsilon,\nu)$ in Ref. \cite{KT} 
for values of $\epsilon$ and $\nu$ for which the latter can be 
translated to the case of shadow processes. Such translation and the 
subsequent comparison of the results was done for the periodic 
instantons. The comparison shows that our perturbative results 
do not differ significantly from the exact ones for $\epsilon \leq 0.25$ 
or, equivalently, for $\nu \leq \nu_{p}(0.25) \approx 0.2$. Thus, the 
intesection of the region $\epsilon \leq 0.25$, $\nu \leq 0.2$ with 
the region for which the saddle point solutions exist 
(Fig.~\ref{fig:region}) can be regarded as a rough estimate of the 
range of validity of the next-to-leading approximation. 

We would like to mention that, actually, in Ref. \cite{KT} the function 
${\cal F}(\epsilon,\nu)$ (see Eq. (\ref{calF-def}))  
was calculated for points $(\epsilon,\nu)$ such that $0.5 < \epsilon < 3$,  
$0.2 < \nu < 1$ and $0 \leq {\cal F}(\epsilon,\nu) \leq 0.6$. For the 
range $0 \leq \epsilon < 0.25$ and $0 \leq \nu < 0.2$ and 
away from the line of periodic instantons methods of Ref. \cite{KT} 
do not allow to obtain the value of ${\cal F}$. Therefore, at the moment our 
perturbative calculations are the only ones which give quantitative 
behaviour of the supression factor in this range.

We should stress that our result for the propagator correction is only 
the leading term of the expansion in powers of $(m\rho)^{2}$ and 
$\rho^{2}/\tau^{2}$. This level of accuracy is determined by the fact 
that in formula (\ref{S-ci}) for the action of the constrained 
instanton terms ${\cal O}(m^{4}\rho^{4})$ were not taken into account. 
In Sect.~\ref{approx} we formulated the procedure of calculation 
of the constraint independent terms of the leading and 
next-to-leading corrections. The procedure essentially relies on the 
condition that the saddle point values for $\trho^{2}$ and 
$\trho^{2}/\ttau^{2}$ are small in the range of $(\epsilon,\nu)$ under 
consideration. We checked that for the numerical saddle point 
solutions found the condition holds true. This conclusion is 
also verified analytically for the regime of very small energies, namely 
when relation (\ref{eps-small}) is valid. This shows the consistency 
of the procedure of calculation. From this discussion it is clear 
that before calculating next-next-to-leading corrections to the 
function $F(\epsilon,\nu)$ one should compute terms ${\cal O}(\trho^{4})$ 
in the leading and propagator corrections.   

We also proved the cancellation of terms singular in the limit
$\nu \rightarrow 0$ in the propagator correction $F^{(2)}(\epsilon,\nu)$ 
in a rather general context. The proof is essentially based on the 
general structure of the $W^{(1)}$ and $W^{(2)}$ terms, the saddle 
point equations and the factorization property of the asymptotics 
of the propagator residue, following from a general formula of 
Ref. \cite{Vo1}.  
As we have explained, the problem of singularities in $\nu$ 
is closely related to the problem
of quasiclassical evaluation of contributions of initial states and
initial-final states. In Sect.~\ref{asympt} we discussed this issue 
within the approach proposed by Mueller in Ref. \cite{Mu92}. 
Namely, we calculated three leading terms of the asymptotics 
of the instanton propagator at large $s$ and showed that with the  
appropriate choice of the propagator constraint the 
$s\ln s$- and $s$-terms cancell out. According to Ref. \cite{Mu92}, with such 
propagator the problem of semiclassical calculation of contributions 
due to initial states and initial-final states can be tackled properly. 

We expect that our results may provide some insight to the understanding 
of the structure of the suppression factor of the cross section of 
instanton induced processes in more realistic models, like QCD and 
the electroweak theory. They may also help to describe some features 
of the behaviour of such cross sections for  
of energies $E \ll E_{sph}$, which are of interest for 
some planned high energy experiments \cite{Ri2}.  

\section*{Acknowledgments}

{\tolerance=500 
We would like to thank A. Ringwald and V. Rubakov for
discussions and valuable comments.  Y.K. acknowledges financial
support from the Russian Foundation for Basic Research (grant
00-02-17679), "University of Russia" grant 990588 and grant
CERN/P/FIS/15196/1999. The work of P.T.is supported in part by the
Swiss Science Foundation, grant 21-58947.99.

\section*{Appendix A}
\def\theequation{A\arabic{equation}}
\setcounter{equation}{0}

Evaluating Eq. (\ref{saddle-W2}) at the saddle point solution 
(\ref{sol-r2}) - (\ref{sol-x}) with the residue of the instanton 
propagator given by expression (\ref{R1}) we obtain the propagator 
correction to the function $F(\epsilon,\nu)$ at $\nu \rightarrow 0$ by 
straightforward calculation. Before presenting the result 
let us introduce some definitions and explain the origin of some 
terms. 

First we write the $s$-variables, defined by Eqs. (\ref{saa}), (\ref{sab}), 
as 
\[
s_{\#}(\bk,\bq) = 2kql_{\#} [ a_{\#}(k,q) - \cos \theta], 
\]
where $\# = aa,ab,bb$, $k$ and $q$ stand for $|\bk|$ and $|\bq|$, 
respectively, $l_{aa}=l_{bb}=-1$, $l_{ab}=1$, and $\theta$ is the 
angle between $\bk$ and $\bq$. While calculating the function 
${\cal S}_{\#}(k,q)$, given by Eq. (\ref{cS-def}), one encounters the 
integrals 
\bea
M_{\#,1}(k,q) & = & \int_{0}^{\pi} \sin \theta d\theta \ln (a_{\#}(k,q) - 
\cos \theta) = a_{\#} \ln \frac{a_{\#}+1}{a_{\#}-1} + \ln (a_{\#}^{2}-1) -2, 
\nonumber \\
M_{\#,2}(k,q) & = & \int_{0}^{\pi} \sin \theta d\theta 
(a_{\#}(k,q) - \cos \theta) \ln (a_{\#}(k,q) - \cos \theta) \nonumber \\
 & = &  
\frac{a_{\#}^{2}+1}{2} \ln \frac{a_{\#}+1}{a_{\#}-1} + 
a_{\#}\ln (a_{\#}^{2}-1) -a_{\#}.  \label{Mgen2}
\eea
The logarithmic terms in the integrands above are due to the 
logarithmic terms in the function $P(\rho^{2}s)$ (see Eq. (\ref{R1})). 
To get contributions to the propagator correction the functions 
$M_{\#,1}(k,q)$ and $M_{\#,2}(k,q)$ are to be integrated over $k$ and $q$ 
(see Eq. (\ref{int-gen1})). Hence, we define 
\[
N_{\#,i}(\tau_{1},\tau_{2}) = \int_{0}^{\infty} \frac{k^{2}dk}{w_{k}} 
\frac{q^{2}dq}{w_{q}} e^{-w_{k}\tau_{1}} e^{-w_{k}\tau_{2}} (kq)^{i-1} 
M_{\#,i}(k,q), 
\]
where $i=1,2$. In particular, we will need the function $N_{ab,2}$ in 
the case when one of its arguments goes to zero. We define 
\[
\eta (\tau) = \frac{\tau_{1}^{2}}{2}
\left. N_{ab,2}(\tau,\tau_{1}) \right|_{\tau_{1} \rightarrow 0}. 
\]
It can be shown that 
\[
\eta (\tau) = \int_{0}^{\infty} \frac{k^{3}dk}{w_{k}} e^{-w_{k}\tau} 
M_{ab,2}(k,\infty), 
\]
where $M_{ab,2}(k,\infty)$ is calculated from (\ref{Mgen2}): 
\beq
M_{ab,2}(k,\infty) = \frac{(w_{k}+k)^{2}}{2k^{2}} \ln \frac{w_{k}+k}{m} 
- \frac{(w_{k}-k)^{2}}{2k^{2}} \ln \frac{w_{k}-k}{m} 
- \frac{w_{k}}{k} \left( 1 + 2\ln \frac{k}{m} \right).  \label{Mab2}
\eeq
The logarithmic terms in the propagator residue $P(\rho^{2}s)$ also 
give rise to the intrgrals of the form 
\[
\int_{0}^{\infty} \frac{dk}{w_{k}} \left( \frac{k}{m} \right)^{n} 
e^{-w_{k}\tau} \ln \frac{k}{m}.
\]
They can be written as 
\[
\frac{d}{d \beta} \left. \int_{0}^{\infty} \frac{dk}{w_{k}} 
\left( \frac{k}{m} \right)^{\beta} e^{-w_{k}\tau} \right|_{\beta = n} 
= \frac{d}{d \beta} \left[ \frac{1}{\sqrt{\pi}} \left( \frac{2}{m\tau} 
\right)^{\beta/2} \Gamma \left( \frac{\beta + 1}{2} \right) 
K_{\beta/2}(m\tau) \right]_{\beta = n}.
\]
Let us introduce the function 
\[
\Psi (m\tau) = \frac{1}{m\tau}\frac{d}{d \beta} K_{\beta}(m\tau)|_{\beta = 1} 
+ \Phi (m\tau) \left[ C_{E} + \ln \frac{m\tau}{2} \right], 
\]
where $\Phi (m\tau)$ is given by Eq. (\ref{Phi-def}). Finally, we define 
the functions 
\bea
{\cal A}_{1}(m\tau) & = & -\frac{1}{2} \Psi'(m\tau) + \frac{1}{m\tau} 
\Phi(m\tau) + \frac{1}{2} \eta (\tau), \nonumber \\
{\cal A}_{2}(m\tau) & = & \Phi (m\tau) \Psi (m\tau) + \frac{1}{2} 
N_{bb,1}(\tau,\tau), \nonumber \\
{\cal A}_{3}(m\tau) & = & \Phi' (m\tau) \Psi' (m\tau) - \frac{2}{m\tau} 
\Phi (m\tau) \Phi' (m\tau) - \Phi (m\tau) \Psi (m\tau) + 
\frac{1}{2} N_{bb,2}(\tau,\tau). \nonumber 
\eea

The partial contributions to the function $F(\epsilon,\nu)$ in 
the limit $\nu \rightarrow 0$ are equal to 
\bea
& & F^{(2)}_{(i-i)} (\epsilon,\nu) = 
-32 \alpha_{1} \upsilon \trho^{6} \frac{\tgamma^{2}}{(\tchi-\ttau)^{6}} 
\left[ \frac{71}{30} + \ln \frac{\trho^{2}}{(\tchi - \ttau)^{2}} \right] 
  \label{F2-ii-2} \\
& & = \frac{8\alpha_{1}}{\upsilon^{2}} 
 \frac{\teps^{3}}{\Phi'(\ttau)} \left[ 2 \ln \frac{1}{\tnu} - 
 12 \Phi (\ttau) + 2\ln \left( -\upsilon \Phi'(\ttau) \right) 
 - 6C_{E} + 4\ln 2 - \frac{109}{30} \right],  \label{F2-ii} \\
& & F^{(2)}_{(i-f)}(\epsilon,\nu) = 
 \frac{32 \alpha_{1} \upsilon \trho^{6} \tgamma}{(\tchi-\ttau)^{3}} 
\left[ - \Phi' (\ttau) \left( \ln \frac{\trho^{2}}{\ttau (\tchi - \ttau)}
+ \frac{71}{30} - \ln 2 \right) + {\cal A}_{1}(\ttau) \right] 
  \label{F2-if-2} \\
& & = -\frac{16\alpha_{1}}{\upsilon^{2}} 
 \frac{\teps^{3}}{\Phi'(\ttau)} \left[\ln \frac{1}{\tnu} -
 8\Phi (\ttau) + \ln \left( -\upsilon \Phi(\ttau) \right) - 
 4C_{E} - \frac{49}{30} + 2\ln 2 \right.  \nonumber \\
 & & \left. -\ln \ttau - 
 \frac{{\cal A}(\ttau)}{\Phi'(\ttau)} \right], \label{F2-if}  \\
& & F^{(2)}_{(f-f)} (\epsilon,0) = 
 - 8 \alpha_{1} \upsilon \trho^{6} 
\left[ \left( \Phi' (\ttau) \right)^{2} 
\left( \ln \frac{\trho^{2}}{\ttau^{2}} + \frac{28}{15} - \ln 2 \right) 
  \right.    \label{F2-ff-2} \\
 & & - \left.
\left( \ln \frac{\trho^{2}}{\ttau^{2}} + \frac{28}{15} - \ln 2 \right)  
\Phi^{2}(\ttau) + {\cal A}_{3}(\ttau) \right] \nonumber  \\
& & - 4 \alpha_{2} \upsilon \trho^{4} \left[ 
\left( \ln \frac{\trho^{2}}{\ttau^{2}} + \frac{73}{15} - \ln 2 \right)  
\Phi^{2}(\ttau) + {\cal A}_{2}(\ttau) \right]  \nonumber \\
& & = \frac{8\alpha_{1}}{\upsilon^{2}} 
 \frac{\teps^{3}}{\Phi'(\ttau)} \left\{ -4 \Phi (\ttau) -2\ln \frac{\ttau}{2} 
-2C_{E} - \frac{2}{15} - \ln 2 + \frac{{\cal A}_{3}
(\ttau)}{(\Phi'(\ttau))^{2}}  \right. \nonumber \\
& & - \left. \frac{\Phi^{2}(\ttau)}{(\Phi'(\ttau))^{2}} \left[ 
- 4 \Phi (\ttau) - \ln \frac{\ttau^{2}}{2} - 2C_{E} -\frac{2}{15} - \ln 2 
\right] \right\} \label{F2-ff} \\ 
& & - \frac{\alpha_{2}}{\upsilon} \teps^{2} 
\frac{4\Phi^{2}(\ttau)}{(\Phi'(\ttau))^{2}} \left[ 
-\ln 2 + \frac{73}{15} - 2C_{E} -2 - 4\Phi(\ttau) \ln \frac{\ttau^{2}}{4} 
+ \frac{{\cal A}_{2}(\ttau)}{\Phi^{2}(\ttau)} \right],  
\eea
where, as before, we use the notation $\upsilon = 192 \pi^{2}$. 
Note that these are the terms $\ln \trho/(\tchi-\ttau)$ in Eq. (\ref{F2-ii-2}) 
and Eq. (\ref{F2-if-2}) that give rise to singular terms $\ln (1/\nu)$. 
The latter are shown explicitly in Eqs. (\ref{F2-ii}), (\ref{F2-if}).  

For $\tau \ll 1$ the expressions above simplify considerably. 
The functions ${\cal A}_{i}(m\tau)$ 
have the properties 
\bea
\frac{{\cal A}_{1}(m\tau)}{\Phi'(\tau)} & = & - \ln 2 + {\cal O}(m\tau), 
\; \; \frac{{\cal A}_{2}(m\tau)}{\Phi^{2}(m\tau)} = \ln 2 -1 + 
{\cal O}(m\tau), \nonumber \\
\frac{{\cal A}_{3}(m\tau)}{(\Phi'(m\tau))^{2}} & = & \ln 2 + \frac{1}{2} + 
{\cal O}(m\tau).   \nonumber 
\eea
Using these relations it is easy to calculate the expressions for 
the partial contributions (\ref{F2-ii}) - (\ref{F2-ff}) 
for $\tau \ll 1$. One gets
\bea
F^{(2)}_{(i-i)}(\epsilon,\nu) & = & - \frac{4\alpha_{1}}{\upsilon^{2}} 
 (\teps \ttau)^{3} 
\left[ 2 \left( 1 + \frac{\ttau^{2}}{4} + {\cal O}(\ttau^{3}) \right) 
\ln \frac{1}{\tnu} - \frac{12}{\ttau^{2}} - 12 \ln \frac{\ttau}{2} 
   \right. \nonumber \\
& + & \left. 2 \ln \upsilon  - 12 C_{E} - \frac{109}{30} \right], 
   \label{F2-ii-1} \\
F^{(2)}_{(i-f)} (\epsilon,\nu)& = & \frac{8\alpha_{1}}{\upsilon^{2}} 
(\teps \ttau)^{3} 
\left[ \left( 1 + \frac{\ttau^{2}}{4} + {\cal O}(\ttau^{3}) \right) 
\ln \frac{1}{\tnu} - \frac{8}{\ttau^{2}} - 8 \ln \frac{\ttau}{2} 
   \right. \nonumber \\
& + & \left.\ln \upsilon - 8C_{E} - \frac{49}{30} \right], 
   \label{F2-if-1} \\
F^{(2)}_{(f-f)} (\epsilon,0)& = & - \frac{4\alpha_{1}}{\upsilon^{2}} (\teps \ttau)^{3} 
\left[ - \frac{4}{\ttau^{2}} - 4 \ln \frac{\ttau}{2} 
 - 4C_{E} + \frac{41}{30} \right]  
   \label{F2-ff-1} \\
& + & \frac{4\alpha_{2}}{\upsilon} \left[ 1 + \frac{\ttau^{2}}{4} 
\left( 8 \ln \frac{\ttau}{2} + 8 C_{E} - \frac{43}{15} \right) 
+{\cal O}(\ttau^{4}) \right]. \nonumber 
\eea

\end{document}